\documentclass[aps,superscriptaddress,twocolumn,pre,longbibliography,floatfix]{revtex4-1}
\usepackage{amsmath,amsthm,amsfonts,amssymb,bm,graphicx,color,mathpazo,times, braket}
\usepackage[colorlinks={true}, citecolor={blue}, filecolor={blue}, linkcolor={red}, urlcolor={blue}]{hyperref}
\usepackage[caption=false]{subfig}
\usepackage{graphicx}
\usepackage{graphicx}
\usepackage{soul}
\usepackage{appendix}
\usepackage{url}
\usepackage{amsmath}

\allowdisplaybreaks

\begin{document}
\title{Enhancing low-temperature quantum thermometry and magnetometry via quadratic interactions in optomechanical-like systems}
\title{Squeezing- and non-Gaussianity-enhanced quantum thermometry and magnetometry via quadratic interactions in optomechanical-like systems}

	\author{Asghar Ullah}
 \email{aullah21@ku.edu.tr}
\affiliation{Department of Physics, Ko\c{c} University, 34450 Sar\i yer, Istanbul, T\"urkiye}

\author{\"Ozg\"ur E. M\"ustecapl\i o\u glu}	
	\email{omustecap@ku.edu.tr}
	\affiliation{Department of Physics, Ko\c{c} University, 34450 Sar\i yer, Istanbul, T\"urkiye}
	\affiliation{T\"UBITAK Research Institute for Fundamental Sciences (TBAE), 41470 Gebze, T\"urkiye}

 \date{\today} 
%**************************************************************%
\begin{abstract}
Standard optomechanical sensors operating in the low-temperature regime often face fundamental precision limits imposed by vacuum fluctuations. Here, we demonstrate that moving beyond conventional radiation-pressure interactions and exploiting quadratic coupling can surpass these limits, generating intrinsic squeezing and non-Gaussian features in the probe state.
We study quantum thermometry and magnetometry in a coupled two-resonator system, focusing on estimating the temperature of a thermal bath and an external magnetic field. The resonators are assumed to be in thermal equilibrium with a common bath, while a weak magnetic field acts on one of the resonators. We perform measurements on a single resonator, which serves as the probe to estimate both parameters. We compute the quantum Fisher information of the probe for two different interaction models between the resonators. Our results show that the counter-rotating terms in the quadratic interaction naturally induce squeezing at intermediate coupling and strong non-Gaussian correlations as the coupling increases further. These effects yield orders-of-magnitude enhancement in sensitivity in the low-temperature and weak-field regimes compared to standard radiation-pressure couplings. Finally, we investigate multiparameter estimation and find that, although the optimal measurements remain compatible, statistical correlations between parameters prevent the simultaneous estimation of temperature and magnetic field from attaining single-parameter precision.

\end{abstract}	
\maketitle
%**************************************************************%
\section{Introduction}
Quantum metrology establishes a systematic framework for exploiting nonclassical resources, including coherence~\cite{Wang_2018,PhysRevA.100.053825,PhysRevLett.107.083601}, entanglement~\cite{Gioannetti2004,PhysRevLett.96.010401,Giovannetti2011,RevModPhys.90.035005}, squeezing~\cite{PhysRevLett.59.2153,PhysRevLett.68.3020,PhysRevLett.130.090802,Polzik1992}, and criticality~\cite{PhysRevA.78.042105,PhysRevLett.99.095701,PhysRevA.78.042105,PhysRevX.8.021022,PhysRevLett.124.120504,PhysRevLett.126.010502,PhysRevLett.129.090503,PhysRevLett.130.240803,PhysRevLett.133.040801,PhysRevLett.133.040801} to achieve measurement sensitivities beyond classical bounds~\cite{RevModPhys.89.035002,RevModPhys.90.035006,MONTENEGRO20251}. Quantum metrology plays a pivotal role in the development of emerging quantum technologies, with key experimental realizations ranging from gravitational-wave detection~\cite{PhysRevA.33.4033,PhysRevA.88.041802,PhysRevD.23.1693,Abadie2011,Aasi2013}, high-resolution spectroscopy~\cite{PhysRevA.46.R6797}, and state-of-the-art atomic clocks~\cite{PhysRevLett.116.063001,King2022,Katori2011,Valencia2004,PhysRevA.72.042301}, to quantum sensors used for searches for elementary particles~\cite{Jiang2021,PRXQuantum.4.020101} and precision measurements of fundamental constants of the Universe~\cite{Rosi2014}.

In recent years, quantum thermometry has attracted significant interest as a powerful tool for the development of quantum technologies~\cite{Dedyulin_2022}. To estimate the temperature of a quantum system, a probe is coupled to the sample and subsequently measured. In equilibrium thermometry, the probe is allowed to thermalize with the system, and temperature information is extracted from energy measurements that maximize the quantum Fisher information (QFI)~\cite{Mehboudi_2019,DePasquale2018,Mehboudi_2015,DePasquale2016,PhysRevA.96.062103,PhysRevA.82.011611,PhysRevLett.114.220405,Mok2021,abiuso2023optimal,Mihailescu_2024}. Alternatively, higher precision can be achieved using transient estimation schemes, in which measurements are performed before the probe reaches thermal equilibrium~\cite{PhysRevA.98.050101,PhysRevA.99.062114,PhysRevResearch.2.033498,Sekatski2022optimal,PhysRevA.91.012331,AncillaD}. Single- and multi-qubit probes are particularly effective within specific temperature ranges~\cite{AncillaD,Campbell_2018,Mok2021,PhysRevA.96.062103}, with further precision enhancements achievable via highly degenerate probes~\cite{PhysRevLett.114.220405,Campbell_2018} or the inclusion of ancillary qubits~\cite{PhysRevResearch.5.043184,Ullah_2025MTCSs,zhang2022}.

Superconducting microwave resonators are widely used in quantum information processing, sensing, and hybrid quantum architectures~\cite{PhysRevA.69.062320,Clarke2008,Devoret2013,Wallraff2004}. In parallel, nonlinear bosonic interactions originally explored within the field of quantum optomechanics have enabled remarkable advances in precision measurement, including near–quantum-limited detection~\cite{Regal2008,Teufel2009,Hertzberg2010}, state transfer~\cite{Palomaki,Palomaki2013}, and the generation of nonclassical states~\cite{Brooks2012,Safavi-Naeini_2013}. Importantly, several of these interaction regimes can be faithfully realized in purely electrical circuits, where nonlinear couplings between microwave resonators give rise to effective optomechanical-type interactions~\cite{RevModPhys.86.1391,PhysRevLett.112.203603,Bertet,PhysRevA.91.033835}. 

%From a metrological perspective, coupled resonators with optomechanical-like interactions provide a natural platform for quantum sensing. Thermometric protocols originally developed for optomechanical systems—such as spectral asymmetry measurements~\cite{Safavi-Naeini_2013,Chowdhury_2019,PhysRevApplied.12.044066} and probe-based estimation schemes~\cite{PhysRevA.84.032105,PhysRevB.88.155409}—can be directly reformulated in terms of resonator–resonator couplings. In this context, thermometry in the nonlinear interaction regime has been shown to enable enhanced precision and saturation of the quantum Fisher information bound~\cite{39bt-37yl,PhysRevResearch.2.043338}, while quantum correlations have been exploited to infer temperature at thermal equilibrium~\cite{science.aag1407}. Furthermore, related hybrid architectures demonstrate how information about system parameters, such as coupling strengths, can be efficiently extracted via an auxiliary mode acting as a quantum probe~\cite{PhysRevResearch.4.033036}. 

Early studies focused on linearized or radiation-pressure-type interactions for thermometry, such as in optomechanical or superconducting resonator systems, where temperature is inferred from spectral asymmetry or from the response of an auxiliary probe mode~\cite{Safavi-Naeini_2013, Chowdhury_2019, PhysRevApplied.12.044066, PhysRevA.84.032105, PhysRevB.88.155409}. In the nonlinear radiation-pressure regime, enhanced precision—and in some cases saturation of the QFI bound—can be achieved, particularly when strong correlations or nonclassical probe states are involved~\cite{39bt-37yl, PhysRevResearch.2.043338, science.aag1407}. In contrast, quadratic interactions contain counter-rotating terms that can induce intrinsic squeezing and non-Gaussian features in the probe mode without the need for external driving~\cite{PhysRevLett.132.053601}. These interaction-induced modifications enhance the structure of the probe state, allow the higher-order correlations to encode information about small parameter variations, and thereby enhance its sensitivity, suggesting a natural advantage for thermometry and magnetometry at low temperatures.

In this work, we adopt a coupled-resonator framework that captures the essential interaction physics while enabling a systematic comparison between quadratic~\cite{PhysRevLett.132.053601} and radiation-pressure-type~\cite{PhysRevA.90.053833} couplings, where the interaction form is used to engineer the probe’s quantum metrological resource for sensing tasks. Rather than targeting the temperature of an individual resonator, we use the coupled-resonator system to estimate the temperature of a thermal bath and a weak magnetic field applied to one of the resonators. We assume that the system is in thermal equilibrium with a low-temperature common bath, while a weak magnetic field acts on a single resonator. We follow a local measurement scheme in which measurements are performed only on the probe resonator. We use the QFI as a figure of merit to quantify the precision with which the unknown parameters encoded in the reduced density matrix of the probe can be estimated. By considering two different interaction Hamiltonians between the resonators, we analyze how the interaction form influences the probe state and, in turn, the achievable QFI for the parameters of interest. We find that quadratic interactions can provide a substantial enhancement in estimation precision compared to radiation-pressure-type couplings, particularly in the low-temperature and weak-field regimes. In the intermediate-coupling regime, they generate intrinsic, equilibrium-induced squeezing that enhances magnetometric sensitivity, while at strong coupling, they produce pronounced non-Gaussian features that significantly improve thermometric precision. We further find that the simultaneous estimation of temperature and magnetic field exhibits a trade-off whose structure depends on the form of the interaction. While quadratic coupling substantially enhances the precision of single-parameter estimation, it introduces statistical correlations between temperature and magnetic-field estimators, degrading multiparameter estimation performance even with squeezing and non-Gaussianity.

The rest of the paper is organized as follows. In Sec.~\ref{model}, we present the model system for quantum metrology. Section~\ref{QET} introduces the key concepts of quantum parameter estimation theory. The results are presented in Sec.~\ref{res}, where we discuss temperature and magnetic-field estimation using QFI and CFI, and then their joint estimation. Finally, we summarize our findings in Sec.~\ref{conc}.

%****************************************************************%
\section{ Model}\label{model}
%****************************************************************%

\begin{figure}[t!]
    \centering
    \includegraphics[scale=0.42]{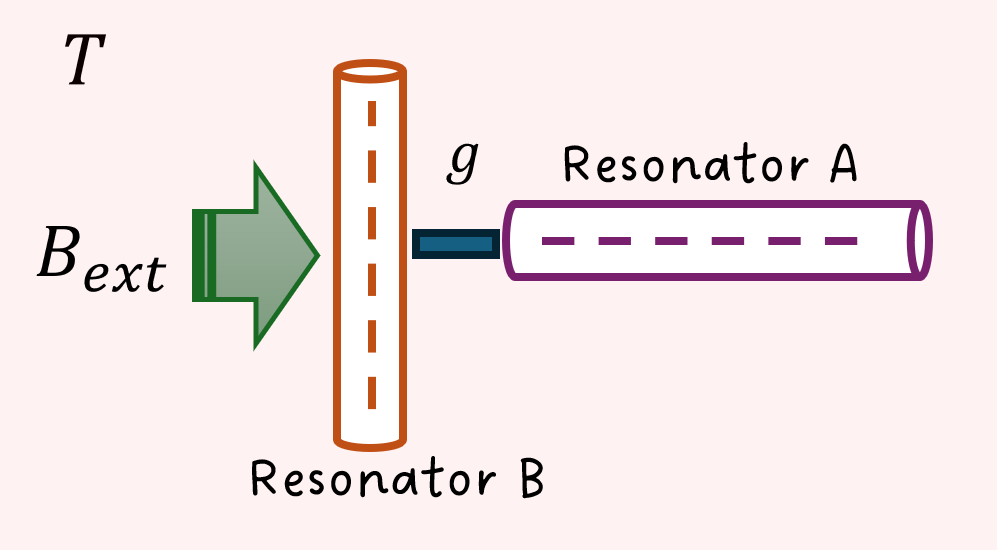}
    \caption{ Schematic of quantum sensing system consisting of resonator~A and resonator~B via an effective optomechanical-like coupling with strength $g$. Both resonators are immersed in a common thermal bath at temperature $T$, while an external magnetic field $B_{\text{ext}}$ is applied to resonator~B.}
    \label{fig:mod}
\end{figure}

We consider a quantum sensing setup consisting of two single-mode resonators, denoted as resonator~A and resonator~B, which are mutually coupled and embedded in a common thermal bath at temperature $T$, as shown in Fig.~\ref{fig:mod}. In addition, resonator~B is subjected to an external magnetic field $B$, which also serves as the parameter to be estimated. 

The total Hamiltonian of the system (setting $\hbar=1$ throughout) can be written as~\cite{PhysRevA.51.2537}
\begin{equation}
\hat{H} = \hat{H}_0 + \hat{H}_B + \hat{H}_I ,
\label{mod}
\end{equation}
where the free Hamiltonian of the two resonators is
\begin{equation}
\hat{H}_0 = \omega_A \hat{a}^\dagger \hat{a} + \omega_B \hat{b}^\dagger \hat{b},
\end{equation}
and the magnetic-field-induced driving of resonator~B is described by
\begin{equation}\label{Magnetic_Field}
\hat{H}_B = B_{\text{ext}} (\hat{b} + \hat{b}^\dagger),
\end{equation}
where resonator~A (B) has a characteristic frequency $\omega_A$ ($\omega_B$) and is described by the bosonic annihilation (creation) operators $\hat{a}\,(\hat{a}^\dagger)$ and $\hat{b}\,(\hat{b}^\dagger)$, respectively, satisfying the canonical commutation relations
$[\hat{a},\hat{a}^\dagger]=\mathbb{I}$, and  $[\hat{b},\hat{b}^\dagger]=\mathbb{I}$.
Here, the external magnetic field effectively couples to the displacement quadrature of resonator~B, inducing a field-dependent shift of its equilibrium position. We emphasize that the Hamiltonian in Eq.~\eqref{Magnetic_Field} considered here is intended as a generic effective two-mode bosonic description rather than a description restricted to a bare optical cavity–mechanical resonator system. Depending on the underlying physical platform, resonator B may represent a mechanical, microwave, superconducting-circuit, magnonic, collective-current, or magnetostrictively active bosonic mode, as realized in a variety of cavity optomechanical, circuit-QED, and cavity-magnomechanical architectures~\cite{PhysRevA.91.033835, PhysRevA.90.053833,Zhang2015,Xufeng2016, RevModPhys.86.1391}. Accordingly, the term $B_\text{ext}(\hat{b}+\hat{b}^\dagger)$ therefore denotes the leading-order linear transduction of an external magnetic field into the displacement-like quadrature of the effective low-frequency mode. In conventional cavity optomechanical realizations, such a coupling would require an intermediary magnetic or magnetostrictive element rather than a direct coupling of a phonon to the magnetic field.
%------------------------------------------------------------------------%
%-------------------------------------------------------------------------%

The interaction between the two resonators is characterized by a coupling strength $g$ and is given by the quadratic interaction~\cite{PhysRevLett.132.053601}
\begin{equation}
\hat{H}_I = g(\hat{a}^\dagger + \hat{a})^2(\hat{b} + \hat{b}^\dagger),
\label{mod1}
\end{equation}
which couples the squared field quadrature of resonator~A to the displacement of resonator~B. Such quadratic interactions arise in hybrid bosonic platforms and are formally analogous to higher-order optomechanical couplings~\cite{PhysRevA.51.2537, PhysRevLett.132.053601}.

In many experimentally relevant regimes, the full quadratic interaction Hamiltonian in Eq.~\eqref{mod1} can be simplified. In particular, under typical operating conditions where resonator~A is weakly excited, the interaction reduces to the standard radiation-pressure–type coupling between the excitation number of resonator~A and the position quadrature of resonator~B. The resulting effective Hamiltonian reads~\cite{Qvarfort2018, RevModPhys.86.1391, PhysRevLett.112.013601}
\begin{equation}
\hat{H}_{\mathrm{eff}} = \omega_A \hat{a}^\dagger \hat{a} + \omega_B \hat{b}^\dagger \hat{b}
+ g\, \hat{a}^\dagger \hat{a} (\hat{b} + \hat{b}^\dagger),
\label{mod2}
\end{equation}
where the interaction term $ g\, \hat{a}^\dagger \hat{a} (\hat{b} + \hat{b}^\dagger) $ couples the excitation number of resonator~A to the displacement of resonator~B. This radiation-pressure interaction coupling plays a central role in quantum sensing and thermometry protocols, as it allows information about the thermal state and external perturbations acting on resonator~B, such as temperature or magnetic fields, to be indirectly encoded in the state of one of the resonators~\cite {Qvarfort2018, RevModPhys.86.1391, PhysRevResearch.2.043338}.

%The two-resonator Hamiltonian (Eq.~\eqref{mod}) considered here is generic and not tied to a specific physical architecture, but can be naturally realized and tailored in several experimental platforms, including circuit quantum electrodynamics and circuit optomechanics~\cite{Massel2011, Massel2012}, as well as silicon-based optomechanical systems~\cite{Fang2017}. Moreover, the effective nonlinear and quadratic couplings required for our quantum thermometry and magnetometry scheme can be engineered using superconducting quantum interference devices (SQUIDs), which enable in situ tunable higher-order interactions between bosonic modes.~\cite{Franco2014}. In this context, regimes in which quadratic interaction terms are retained can have their effective strength engineered in superconducting circuit platforms. Such systems can enable controlled suppression of linear radiation-pressure contributions and enhancement of higher-order couplings, allowing the quadratic interaction to act as an intrinsic equilibrium-generated squeezing and non-Gaussianity as a metrological resource.

The two-resonator Hamiltonians (Eqs.~\eqref{mod1} and \eqref{mod2}) are particularly relevant to superconducting circuit optomechanical architectures, where nonlinear couplings between microwave resonators can be engineered using SQUID-mediated interactions~\cite{PhysRevA.90.053833}. In this architecture, two superconducting resonators are coupled through a SQUID terminating one of the resonators, giving rise to a tunable optomechanical-like interaction between the field quadrature of one resonator and the excitation number of the other. Such superconducting platforms can access strong single-photon coupling regimes and allow in situ tunability of higher-order nonlinear interactions. Motivated by these superconducting resonator systems, we consider the generalized quadratic interaction in Eq.~\eqref{mod1} (see Appendix~\ref{Eq.(4)} for more details), which retains the counter-rotating
two-photon processes beyond the conventional radiation-pressure approximation. As discussed in
Appendix~\ref{Eq.(4)}, the effective Hamiltonian in Eq.~\eqref{mod2} emerges in the weak-coupling regime after neglecting the rapidly oscillating counter-rotating contributions within a rotating-wave-type approximation. Although our analysis is primarily motivated by superconducting circuit implementations, similar higher-order bosonic interactions can also be realized in circuit quantum electrodynamics and circuit optomechanics~\cite{Massel2011, Massel2012, RevModPhys.86.1391}, as well as silicon-based optomechanical systems~\cite{Fang2017}. The coupling constant \(g\) in the present model is an effective nonlinear bosonic coupling and should not be identified directly with the bare dipolar coupling used to define the ultrastrong- or deep-strong-coupling regimes of cavity QED. Consequently, the additional self-energy and higher-order terms required by a microscopic description depend on the specific implementation. Nevertheless, when the effective nonlinear interaction becomes comparable to the low-frequency scale, higher-order corrections inherited from the microscopic platform may quantitatively renormalize the spectrum and the optimal sensing point. We therefore examine the robustness of our conclusions under representative stabilizing and cross-nonlinear corrections in Appendix~\ref{Robustness}.
%----------------------------------------------------------------------%
\begin{figure*}[t!]
    \centering
    \subfloat[]{
    \includegraphics[scale=0.45]{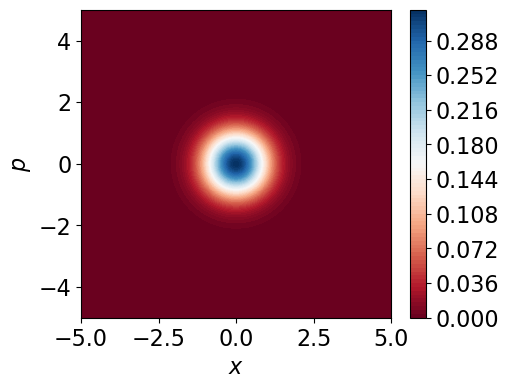}}
    \subfloat[]{
        \includegraphics[scale=0.45]{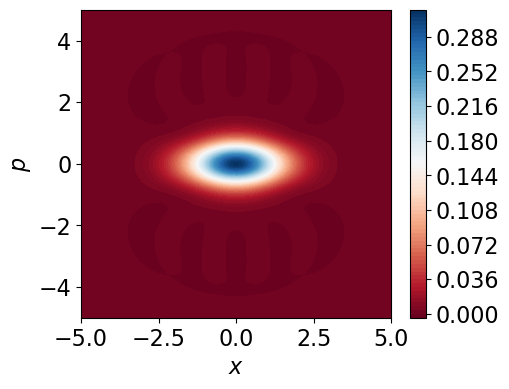}}
        \subfloat[]{
        \includegraphics[scale=0.45]{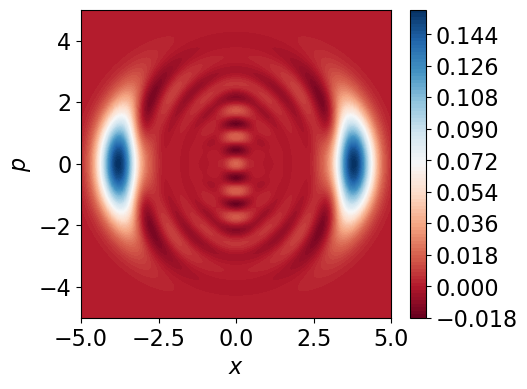}}
        
    \caption{ (a) Wigner function of the probe (resonator~A) obtained under the radiation-pressure interaction ~\eqref{mod2} for coupling strength $g=0.08$. (b)–(c) Wigner functions of the probe under the quadratic interaction Hamiltonian~\eqref{mod1} for coupling strengths $g=0.04$ and $g=0.08$, respectively. The remaining parameters are set to $\omega_A=1$, $\omega_B=0.04$, $T=0.01$, and $B_{\text{ext}}=0.06$. For the quadratic interaction Hamiltonian~\eqref{mod2}, squeezing in the probe state is observed, and when the coupling strength $g$ is increased, it forms two symmetric lobes with interference fringes, whereas no squeezing is generated when the radiation pressure interaction~\eqref{mod2} is considered. 
}
    \label{fig:wf}
\end{figure*}
%----------------------------------------------------------------------%

%*****************************************************************%
\section{Multiparameter quantum estimation theory}\label{QET}
The task of multiparameter quantum estimation is to determine several unknown variables that are collectively encoded in the state of a quantum system. We denote the vector of parameters by $\boldsymbol{\theta} = (\theta_1, \theta_2, \ldots, \theta_d)^T,$
, which is inferred from measurements performed on a parameter-dependent quantum state
\(\rho(\boldsymbol{\theta})\). Information about the parameters is imprinted on the probe through its dependence on \(\boldsymbol{\theta}\), and suitable measurement strategies are employed to extract this information with optimal precision. For a given positive operator-valued measure (POVM) \(\{\Pi_x\}\), the precision of an unbiased estimator \(\tilde{\boldsymbol{\theta}}\) obtained from \(m\) independent experimental runs is bounded by the multiparameter Cramér-Rao inequality~\cite{Helstrom1969,cramer1999mathematical,PhysRevLett.72.3439,paris2009,HELSTROM1967101},
\begin{equation}
\mathrm{Cov}(\tilde{\boldsymbol{\theta}})
\ge \frac{1}{m}\,\boldsymbol{F}^{-1},
\label{eq:multi_crb}
\end{equation}
where \(\mathrm{Cov}(\tilde{\boldsymbol{\theta}})\) is the covariance matrix of the estimator and \(\boldsymbol{F}\) is the classical Fisher information matrix (CFIM). Both matrices are real, symmetric, and positive semidefinite, and the inequality is understood in the sense of matrix ordering.

The elements of the covariance matrix are defined as
\begin{equation}
\mathrm{Cov}(\theta_\mu,\theta_\nu)
= \big\langle
(\theta_\mu - \langle \theta_\mu \rangle)
(\theta_\nu - \langle \theta_\nu \rangle)
\big\rangle,
\end{equation}
with diagonal entries corresponding to the variances
\(\mathrm{Var}(\theta_\mu) = \mathrm{Cov}(\theta_\mu,\theta_\mu)\).
For a fixed measurement basis, the CFIM takes the form~\cite{paris2009}
\begin{equation}
\boldsymbol{F}_{\mu\nu}
=
\sum_x
\frac{
\partial_{\theta_\mu} p(x|\boldsymbol{\theta})\,
\partial_{\theta_\nu} p(x|\boldsymbol{\theta})
}{
p(x|\boldsymbol{\theta})
},
\end{equation}
where
\(p(x|\boldsymbol{\theta}) = \mathrm{Tr}[\rho(\boldsymbol{\theta})\Pi_x]\)
is the probability of obtaining outcome \(x\) for parameters $\boldsymbol{\theta}$.

Maximization of the CFIM over all possible POVMs yields the QFI matrix (QFIM), which sets the ultimate quantum limit to multiparameter estimation. The QFIM can be expressed in terms of the symmetric logarithmic derivative (SLD) operators \(\{L_\mu\}\) as
\begin{equation}
\boldsymbol{\mathcal{F}}_{\mu\nu}
= \frac{1}{2}
\mathrm{Tr} \!\left[ \rho(\boldsymbol{\theta})
\left( L_\mu L_\nu + L_\nu L_\mu\right) \right].
\label{eq:qfim_def}
\end{equation}
The SLD operator \(L_\mu\) corresponding to parameter \(\theta_\mu\) is defined implicitly by the Lyapunov equation
\begin{equation}
\partial_{\theta_\mu}\rho(\boldsymbol{\theta}) =
\frac{1}{2} \left(
L_\mu \rho(\boldsymbol{\theta}) + \rho(\boldsymbol{\theta}) L_\mu
\right).
\end{equation}
By construction, the QFIM is a real, symmetric, and positive semidefinite matrix. Its diagonal elements reduce to the single-parameter QFI, while the off-diagonal elements quantify statistical correlations between different parameters, capturing potential trade-offs in their simultaneous estimation. Using the spectral decomposition $\rho(\boldsymbol{\theta}) =\sum_n q_n \, |\phi_n\rangle\langle\phi_n|$, the elements of the QFIM can be written in a convenient form~\cite{Liu_2020}
\begin{equation}
\boldsymbol{\mathcal{F}}_{\mu\nu} =
\sum_{m,n} \frac{ 2\,\mathrm{Re} \!\left[
\langle \phi_m | \partial_{\theta_{\mu}}\rho(\boldsymbol{\theta}) | \phi_n \rangle
\langle \phi_n | \partial_{\theta_{\nu}}\rho(\boldsymbol{\theta}) | \phi_m \rangle
\right]}{q_m + q_n},
\label{eq:qfim_spectral}
\end{equation}
where $q_m + q_n \neq 0 $. When the SLD operators associated with different parameters are mutually compatible, the ultimate precision limit for simultaneous estimation is set by the quantum Cramér-Rao bound (QCRB). In this case, the covariance matrix of any unbiased estimator \(\tilde{\boldsymbol{\theta}}\) satisfies the matrix inequality~\cite{Helstrom1969,paris2009,holevo2011probabilistic,HELSTROM1967101}
\begin{equation}
\mathrm{Cov}(\tilde{\boldsymbol{\theta}})
\ge
\frac{1}{m}\mathbf{F}^{-1}
\ge
\frac{1}{m}\boldsymbol{\mathcal{F}}^{-1}.
\label{eq:qcrb_multi}
\end{equation}
where \(\boldsymbol{F}\) denotes the CFIM corresponding to a specific measurement strategy, and \(\boldsymbol{\mathcal{F}}\) is the QFIM. The QCRB represents an asymptotic limit that can be approached only under idealized conditions. In particular, its attainability depends on the existence of compatible measurement operators and on the ability to construct estimators tailored to the local structure of the parameter space. 
%-------------------------------------------------------------------%
\section{Results}\label{res}
%-------------------------------------------------------------------%
 
 In this section, we present our results for the individual and joint estimation of the bath temperature $T$ and the external magnetic field $B_{\text{ext}}$ using the coupled two-resonator system described earlier. All numerical simulations are performed using the open-source toolbox QuTiP, which is specifically designed to solve open-quantum-system dynamics~\cite{JOHANSSON20121760}. In addition, all system parameters are rescaled with respect to the characteristic frequency of resonator~A, which we set to $\omega_A=1$, rendering all quantities dimensionless. Throughout our analysis, we assume that the frequency of resonator~B is significantly smaller than that of resonator~A, i.e., $\omega_A \gg \omega_B$ ~\cite{PhysRevA.90.053833}. This hierarchy of energy scales is commonly encountered in hybrid bosonic systems and mirrors the separation of frequencies typically assumed in optomechanical and circuit-based platforms used for sensing~\cite{PhysRevResearch.4.033036, Pirkkalainen2015, PhysRevA.90.053833, Schmidt2020}. Related nonlinear and quadratic optomechanical interactions have also been investigated in cavity optomechanical systems~\cite{Khorasani2018,Sala2018}.
  
For concreteness, one may associate the dimensionless parameters used in this work with experimentally relevant microwave-frequency scales by choosing $\omega_A/2\pi = 10~\mathrm{GHz}$, which is typical for superconducting circuit and circuit-QED platforms~\cite{PhysRevLett.120.227702, Bothner2021}. With this choice, the ratio $\omega_B/\omega_A = 0.04$ corresponds to $\omega_B/2\pi \sim 400~\mathrm{MHz}$, while the coupling range $g/\omega_A \in [0.02,0.08]$ corresponds to coupling strengths $g/2\pi \sim 200\text{--}800~\mathrm{MHz}$, lying within experimentally accessible strong- and ultrastrong-coupling regimes demonstrated in superconducting circuit-QED platforms~\cite{PhysRevB.93.214501,PhysRevB.95.224515,Niemczyk2010}. Moreover, the dimensionless magnetic-field values considered here, typically $B_{\text{ext}}\sim0.2$ and up to $B_{\text{ext}}\sim0.6$ in some cases, correspond to physical magnetic fields of approximately $0.07\text{-}0.21~\mathrm{T}$~\cite{PhysRevApplied.5.044004}.
%------------------------------------------------------------------%
\begin{figure*}[t!]
    \centering
\subfloat[]{
        \includegraphics[scale=0.48]{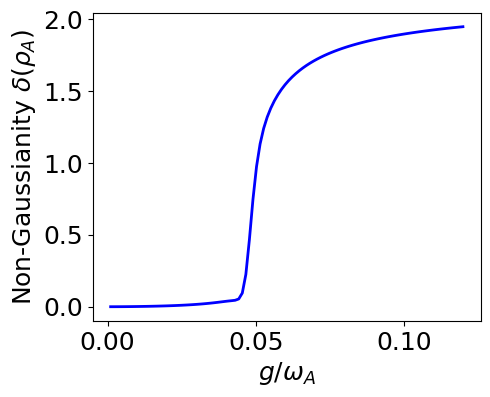}}
        \subfloat[]{
        \includegraphics[scale=0.48]{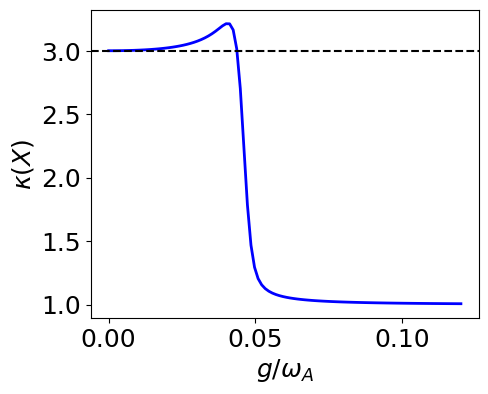}}
        \subfloat[]{
        \includegraphics[scale=0.48]{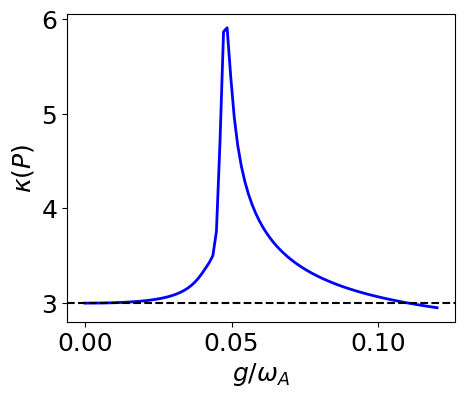}}
    \caption{Panels (d)–(f) display the non-Gaussianity $\delta$ and the quadrature kurtoses $\kappa(X)$ and $\kappa(P)$ of the probe state as functions of the coupling strength $g$ for the quadratic interaction. The parameters are fixed at $\omega_A=1$, $\omega_B=0.04$, $T=0.08$, and $B=0.06$.}
    \label{fig:nG}
\end{figure*}
%------------------------------------------------------------------%

We consider a scenario in which the composite system, described by the total Hamiltonian $\hat{H}$, is weakly coupled to a common thermal bath at temperature $T$ and subjected to a weak external field $B_{\text{ext}}$. The interaction between each resonator and the bath is assumed to be linear in the resonator position and to satisfy the weak-coupling and Markovian approximations~\cite{Breuer}. In this weak system–bath coupling regime and assuming sufficiently long equilibration times, the steady state of the composite system is given by the global Gibbs thermal state~\cite{Breuer,lidar2020lecturenotestheoryopen},
\begin{equation}
\hat{\rho}_T = \frac{e^{-\hat{H}/T}}{\mathcal{Z}},
\label{gibbs}
\end{equation}
where $\hat{H}$ includes the full interaction between the two resonators, $T$ denotes the bath temperature to be estimated, and we have set the Boltzmann constant to unity, $k_B=1$. The corresponding partition function is defined as
\begin{equation}
\mathcal{Z} = \mathrm{Tr}\!\left[e^{-\hat{H}/T}\right].
\end{equation}
Within this framework, the bath temperature $T$ and the magnetic field strength $B_{\text{ext}}$ are globally encoded in the thermal state of the interacting resonators, while the estimation of both parameters is carried out exclusively via local measurements on resonator~A, which we use as a probe. The assumption of sufficiently long equilibration times implies relaxation to the steady state on timescales set by the inverse dissipation rate $\gamma^{-1}$, i.e., $t \gg 1/\gamma$, with practical convergence to the thermal state typically can be achieved for $t \sim 5/\gamma$--$10/\gamma$~\cite{Breuer}.

\subsection{Wigner-function characterization of the probe state}

To characterize the quantum probe in phase space, we employ the Wigner quasiprobability distribution, which provides a complete representation of continuous-variable quantum states. The Wigner function establishes a one-to-one correspondence with the density operator, enabling a phase-space characterization of quadrature fluctuations, squeezing, and non-Gaussian features in bosonic modes.

For a density operator $\hat{\rho}$, the Wigner function in position-momentum space is defined as~\cite{PhysRev.40.749}
\begin{equation}
W(x,p) = \frac{1}{\pi \hbar} \int_{-\infty}^{\infty} dy \;
e^{2 i p y / \hbar}
\left\langle x - y \middle| \hat{\rho} \middle| x + y \right\rangle.
\end{equation}

The Wigner function is real-valued and normalized,
\begin{equation}
\int dx \, dp \; W(x,p) = 1.
\end{equation}
Although it resembles a classical probability distribution, it can take negative values, which serve as a hallmark of nonclassical behavior. This phase-space representation provides a natural framework for analyzing the thermal probe states and their evolution under the interaction Hamiltonians in Eqs.~\eqref{mod1} and~\eqref{mod2}.

Throughout this study, we fix $\omega_A=1$ and $\omega_B=0.04$ and consider two different interaction Hamiltonians given in Eqs.~\eqref{mod1} and~\eqref{mod2} and analyze how these modify the state of the probe. To this end, we plot the Wigner function of the probe state for these two interaction Hamiltonians, as shown in Fig.~\ref{fig:wf}, plotted against the phase space variables $\hat{x}$ and $\hat{p}$. As can be seen from the Wigner functions, the probe is in a thermal state for the interaction Hamiltonian~(\ref{mod2}) as shown in~\ref{fig:wf}(a). For the interaction Hamiltonian~(\ref{mod1}), the probe state exhibits squeezing even for moderate coupling, with the squeezing primarily appearing in the momentum quadrature. In the weak-coupling regime, the probe state remains approximately thermal for both interaction models. As the coupling strength increases, squeezing becomes more pronounced. For example, at $g=0.06$, the Wigner function shows clear squeezing in $\hat{p}$, while at $g=0.08$ (see Fig.~\ref{fig:wf}(c)), the Wigner function develops two well-separated coherent lobes accompanied by pronounced interference fringes and negative regions, indicating strong non-Gaussian features in the probe state. Although radiation-pressure interactions are inherently nonlinear, they remain Gaussian and do not generate non-Gaussianity in thermal equilibrium in the absence of external driving. In contrast, the quadratic interaction considered here includes counter-rotating terms that can generate equilibrium squeezing at intermediate coupling strengths and drive the probe into a strongly non-Gaussian state at larger interaction strengths. It is therefore essential to analyze the structure of the probe state under quadratic interaction in order to identify the parameter regimes where genuine quadrature squeezing occurs and where the physics is instead dominated by non-Gaussian features.

\subsection{Non-Gaussianity and higher-order moment analysis}

To further characterize the probe state beyond its phase-space representation, we quantify its deviation from Gaussianity and analyze higher-order statistical moments of the quadratures. While Gaussian states are completely determined by first and second moments, the presence of interference fringes and Wigner negativity in Fig.~\ref{fig:wf}(c) suggests the emergence of genuinely non-Gaussian features.

We quantify the non-Gaussianity of the probe state using the relative-entropy distance between the probe state $\rho_A$ and the reference Gaussian state $\rho_G$, which has the same first and second moments and is given by~\cite{PhysRevA.78.060303}
\begin{equation}
\delta(\rho_A) = S(\rho_G) - S(\rho_A),
\end{equation}
where $S(\rho_A) = -\mathrm{Tr}(\rho_A \log \rho_A)$ is the von Neumann entropy.
By construction, $\delta(\rho_A)=0$ for Gaussian states and $\delta(\rho_A)>0$ indicates genuine non-Gaussian character. Figure~\ref{fig:nG}(a) shows non-Gaussianity $\delta(\rho_A)$ as a function of $g$ for the probe state. It is obvious that $\delta(\rho_A)$ increases sharply beyond a critical coupling strength $g$, signaling a transition from a nearly Gaussian regime to a strongly non-Gaussian one. Hence, the probe state shows non-Gaussian features in certain parameter regimes. A covariance matrix analysis shows that, although squeezing appears at intermediate coupling ($g \lesssim 0.051$), the minimum rotated quadrature variance in the strong-coupling regime remains above the vacuum limit with vanishing covariance between $X$ and $P$. Hence, the highly non-Gaussian regime is not accompanied by residual squeezing but instead corresponds to a bimodal, cat-like state.

To capture higher-order fluctuations and reveal non-Gaussian characteristics, we also compute the kurtosis of the quadratures
\begin{equation}
X = \frac{\hat{a} + \hat{a}^\dagger}{\sqrt{2}},
\qquad
P = \frac{\hat{a} - \hat{a}^\dagger}{i\sqrt{2}},
\end{equation}
which is defined as
\begin{equation}
\kappa(X) = \frac{\langle (\Delta X)^4 \rangle}{\langle (\Delta X)^2 \rangle^2},
\qquad
\kappa(P) = \frac{\langle (\Delta P)^4 \rangle}{\langle (\Delta P)^2 \rangle^2}.
\end{equation}

For any Gaussian state, the kurtosis satisfies $\kappa(X) = \kappa(P) = 3$ and deviations from this value therefore provide a direct signature of non-Gaussian statistics.  Figure~\ref{fig:nG}(b) shows the kurtosis $\kappa(X)$ in the position quadrature $X$. One can see that as the coupling strength increases, $\kappa(X)$ drops below the Gaussian value and reaches $\kappa(X)< 3$. This sub-Gaussian (platykurtic) behavior indicates a flattened distribution along the $X$ direction. This shows the formation of two well-separated lobes in phase space, as observed in the Wigner function (Fig.~\ref{fig:wf}(c)). The decrease of $\kappa(X)$ therefore signals a redistribution of probability weight away from the center, characteristic of a bimodal-like structure.
%--------------------------------------------------------------------%
\begin{figure}[t!]
    \centering
    \includegraphics[scale=0.75]{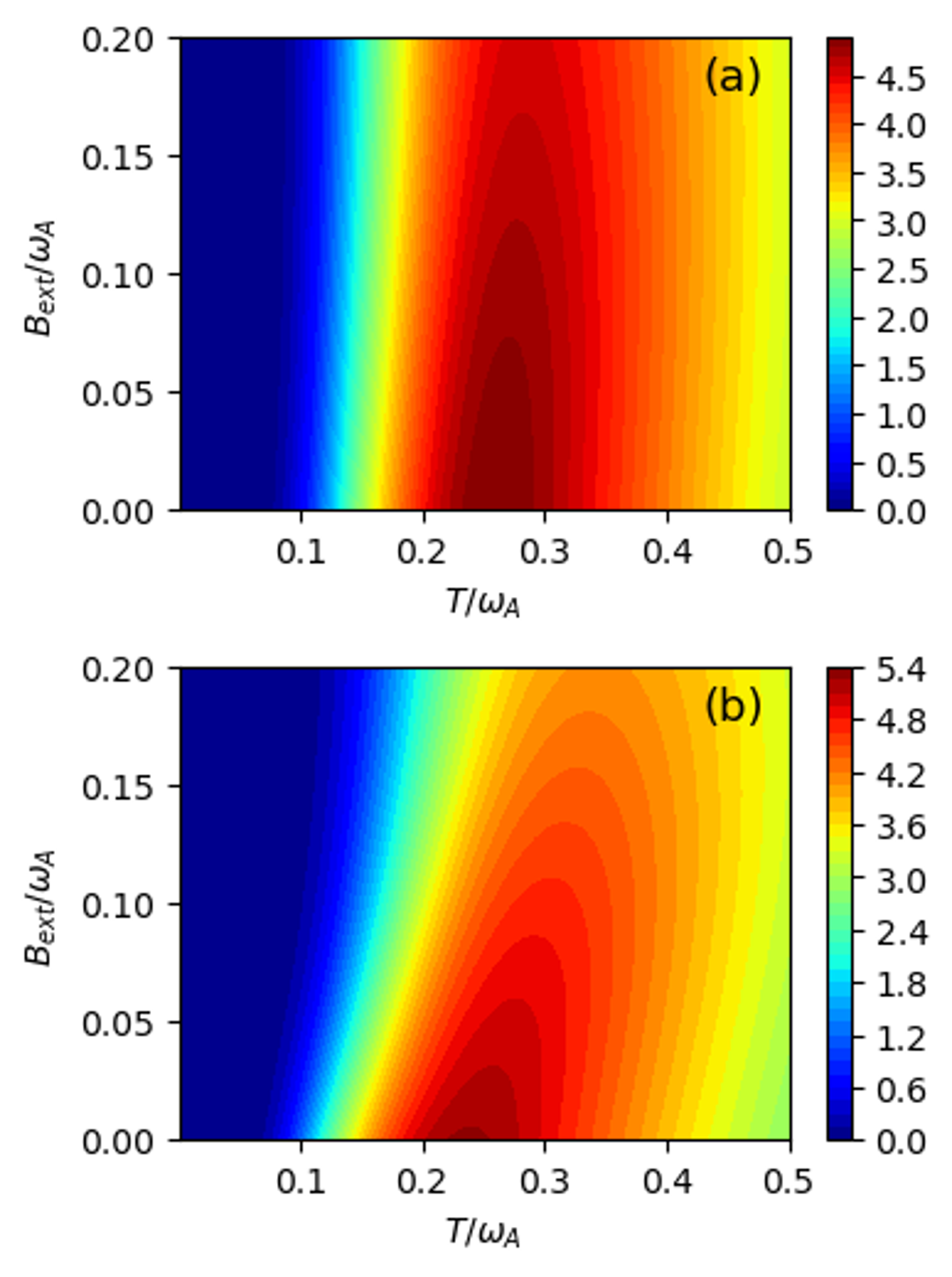}
    \caption{ QFI as a function of the temperature $T$ and magnetic field $B_{\text{ext}}$, obtained using Resonator~A as the local probe under the effective interaction Hamiltonian~\eqref{mod2}. The parameters are set to $\omega_A=1$ and $\omega_B=0.04$. Panels (a) and (b) correspond to coupling strengths $g=0.02$ and $g=0.08$, respectively.
}
 \label{fig2}
\end{figure}
%--------------------------------------------------------------------%
%--------------------------------------------------------------------%
\begin{figure*}[t!]
    \centering
    \includegraphics[scale=0.44]{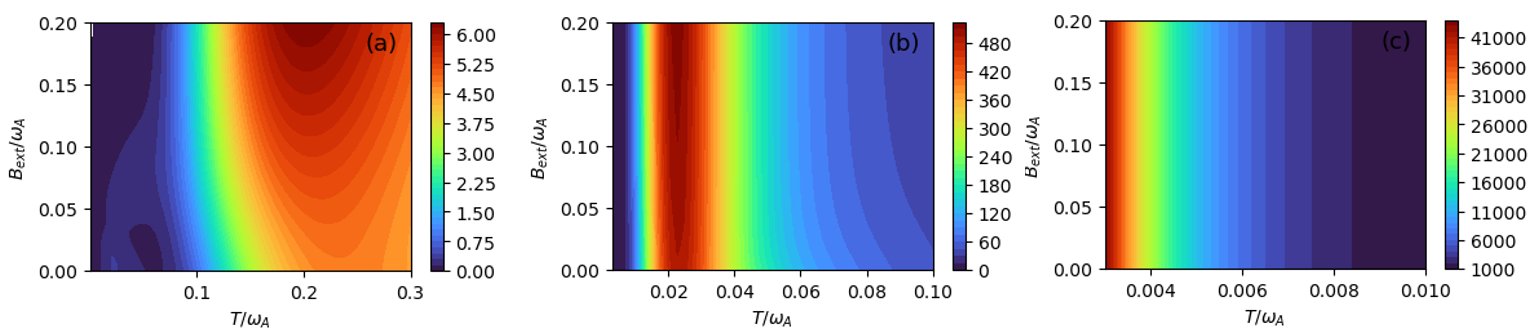}
    \caption{ QFI as a function of the bath temperature $T$ and the external magnetic field $B_{\text{ext}}$, evaluated using using resonator~A as the local probe under the quadratic interaction Hamiltonian~\eqref{mod1}. The other parameters are set to $\omega_A=1$ and $\omega_B=0.04$. Panels (a)-(c) correspond to coupling strengths $g=0.02$, $g=0.06$, and $g=0.08$, respectively.
}
    \label{fig3}
\end{figure*}
%--------------------------------------------------------------------%
In contrast, the kurtosis $\kappa(P)$ of the momentum quadrature $P$ exhibits the opposite trend as shown in Fig.~\ref{fig:nG}(c). Near the onset of coupling $g\approx0.04$, $\kappa(P)$ increases above the Gaussian value and displays a pronounced peak with $\kappa(P) > 3$. This leptokurtic behavior reflects heavy-tailed statistics and enhanced higher-order fluctuations along the $P$ direction. Such behavior is consistent with the emergence of oscillatory interference fringes in the Wigner representation. Importantly, both $\kappa(X)< 3$ and $\kappa(P) > 3$ represent clear deviations from Gaussian statistics. The opposite behavior of the two quadratures highlights the anisotropic deformation of the phase-space distribution induced by the nonlinear interaction, providing further evidence of genuine non-Gaussianity in the probe state.

The quadratic couplings, therefore, provide an intrinsic equilibrium-generated metrological resource that transitions from Gaussian squeezing at moderate interaction strengths to strong non-Gaussianity in the strong coupling regimes. These probe features offer a more direct route to quantum sensing enhanced by non-Gaussianity in the low-temperature and weak field regimes. In the subsequent section, we investigate the effect of this interaction-induced squeezing and non-Gaussianity on the precision of temperature and magnetic-field estimation.

\subsection{Estimation of temperature}

We first consider the estimation of the bath temperature $T$ using resonator~A as a probe. In this context, we analyze the precision of temperature estimation when the interaction Hamiltonian~\eqref{mod2} is considered. We first plot the QFI for the probe state as a function of the temperature $T$ and magnetic field $B$ in Fig.~\ref{fig2}(a) while assuming that the magnetic field $B_{\text{ext}}$ is known. In the weak-coupling regime, such as $g=0.02$, we observe that the QFI is slightly smaller than that of a single harmonic oscillator in thermal equilibrium. However, when $g=0.08$ [cf. Fig.~\ref{fig2}(b)], the maximum value of the QFI increases, although this enhancement is not significant. It is worth noting that in the weak-coupling regime, where $g$ is much smaller than the characteristic resonator frequencies and $\omega_A \gg \omega_B$, the rotating-wave approximation (RWA) is expected to be valid. In this limit, both interaction models lead to thermal-like probe states with no appreciable squeezing and non-Gaussian features, consistent with standard optomechanical expectations. As the coupling strength increases, the quadratic interaction activates counter-rotating terms that are suppressed within the RWA, leading to pronounced beyond-RWA effects and the emergence of intrinsic squeezing and non-Gaussian features at thermal equilibrium (see Fig.~\ref{fig:wf}(d)). This beyond-RWA behavior underlies the enhanced metrological performance, which varies as the coupling strength is varied. We also observe that a strong magnetic field is not advantageous for temperature estimation. The maximum value of the QFI is obtained in the weak magnetic-field region (dark red).

\begin{figure}[t!]
    \centering
    \includegraphics[scale=0.66]{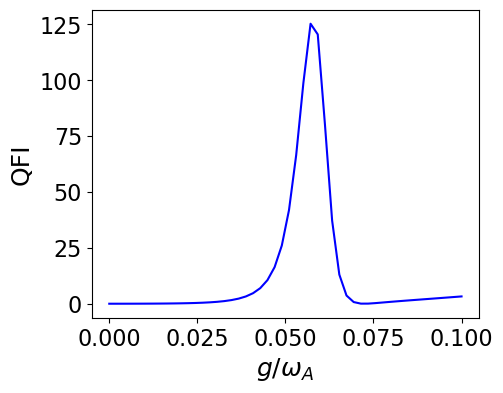}
    \caption{ QFI as a function of coupling strength $g$ for estimation of temperature $T$ using Eq.~\eqref{mod1}. The rest of the parameters are set to $\omega_A=1$, $\omega_B=0.04$, $T=0.06$, and $B_\text{ext}=0.04$.}
    \label{fig:QFI_vs_g}
\end{figure}

To enhance the QFI and thereby improve the precision of temperature estimation, we next consider the quadratic interaction Hamiltonian given in Eq.~\eqref{mod1}. As the state of the probe is now modified by the quadratic interaction, leading to the emergence of squeezing and non-Gaussianity (see Fig.~\ref{fig:wf}(b) and (d), respectively), well-known resources in quantum metrology~\cite{62ks-19fs,PhysRevLett.134.180801, santos2026, PhysRevA.109.052604, Gessner2020}. We plot the QFI in Fig.~\ref{fig3} as a function of $T$ and $B_{\text{ext}}$ for different values of $g$. Figure~\ref{fig3}(a) shows that the QFI for $g=0.02$ is slightly larger than the corresponding QFI in Fig.~\ref{fig2}. However, the maximum value of the QFI now lies in the region of a strong magnetic field $B_{\text{ext}}$. This behavior arises because, at smaller coupling strengths, the state remains close to the Gaussian and genuine quadrature squeezing is present, as indicated by the Wigner function in Fig.~\ref{fig:wf}(b), but the deviation from Gaussianity is still weak (Fig.~\ref{fig:wf}(d)) and does not substantially enhance the estimation precision. However, when the coupling strength increases to $g=0.06$ or $g=0.08$, the state becomes strongly non-Gaussian, while quadrature squeezing disappears, leading to a few orders of magnitude enhancement in the QFI, depicted in Fig.~\ref{fig3}(c). Figure~\ref{fig:QFI_vs_g} shows the QFI as a function of the normalized coupling strength $g/\omega_A$. The QFI exhibits a pronounced nonmonotonic behavior, reaching a maximum at intermediate values of the $g$ and follows a quite similar pattern as shown by non-Gaussianity in Fig.~\ref{fig:wf}(d) and kurtosis in Fig.~\ref{fig:wf}(f). This feature is specific to the quadratic interaction, for which counter-rotating terms become increasingly relevant as $g/\omega_A$ grows. In the weak-coupling regime, where the rotating-wave approximation is effectively valid, the probe state remains thermal-like and Gaussian, resulting in a small QFI. As the coupling strength increases, beyond-RWA effects first generate genuine quadrature squeezing and moderate non-Gaussianity, leading to an enhancement of the QFI. At larger values of $g/\omega_A$, the probe state undergoes a transition to a strongly non-Gaussian regime in which squeezing disappears, and the QFI decreases again, indicating the existence of an optimal quadratic-coupling window for metrological performance. Notably, the QFI remains maximal over a broad range of magnetic field values $B_{\text{ext}}$ and shifts toward lower temperatures as $g$ increases.

This behavior originates from the counter-rotating terms in the quadratic interaction, which act as an effective parametric drive. At moderate coupling, they induce equilibrium squeezing, while at stronger coupling, they drive the probe into a highly non-Gaussian, cat-like state. Unlike radiation-pressure interactions, which rely on thermal fluctuations to encode temperature information, this intrinsic non-Gaussian mechanism persists even as $T \to 0$, enabling higher-order correlations to enhance the susceptibility of the probe to small temperature variations. Consequently, quadratic interactions allow improved thermometric precision in the low-temperature regime compared to radiation-pressure couplings.

In our scheme, squeezing at intermediate coupling and strong non-Gaussianity at larger coupling strengths provide complementary metrological advantages: the former enhances Gaussian sensitivity, while the latter enables temperature probing in regimes where the QFI of the effective Hamiltonian~\eqref{mod2} would otherwise become vanishingly small. The distinct metrological performance of the Hamiltonians~\eqref{mod1} and ~\eqref{mod2} originates from their fundamentally different operator structures. The radiation-pressure interaction is diagonal in the Fock basis of resonator~A and primarily induces number-conditioned displacements of resonator~B, without introducing pair-creation or pair-annihilation processes. In contrast, the quadratic interaction, $(\hat{a}+\hat{a}^{\dagger})^{2}$, contains counter-rotating two-photon terms that couple Fock states differing by two excitations. At moderate coupling strengths, these processes generate quadrature squeezing, whereas stronger interactions give rise to higher-order coherences and non-Gaussian phase-space structures. Simultaneously, the quadratic interaction reshapes the low-energy spectrum, enhancing the sensitivity of the reduced thermal state to variations in temperature and magnetic field. Consequently, the improved QFI arises from the combined effects of interaction-induced spectral restructuring and phase-space deformation, while squeezing and non-Gaussianity should be regarded as observable manifestations of this underlying mechanism rather than independent metrological resources.

%----------------------------------------------------------%
%----------------------------------------------------------%
%************************************************************************%
\subsection{ Estimation of magnetic field} 
%************************************************************************%

\begin{figure}[t!]
    \centering
    \includegraphics[scale=1.1]{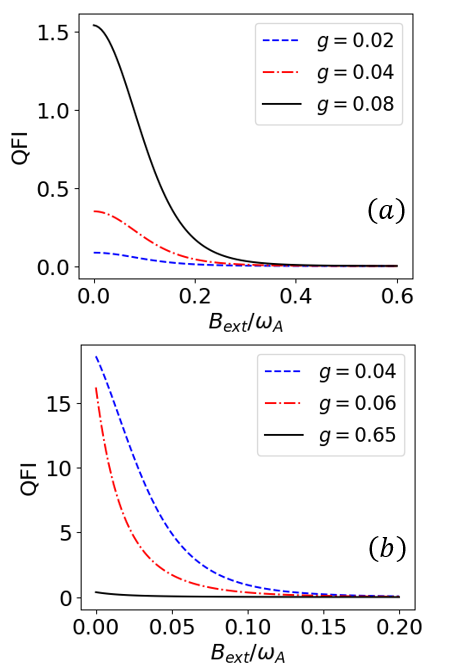}
    \caption{ QFI as a function of the magnetic field $B_{\text{ext}}$ for the estimation of $B$ is shown for different values of the coupling strength. Panel (a) corresponds to the radiation-pressure interaction, while panel (b) corresponds to the quadratic interaction. The other parameters are set to $\omega_A=1$, $\omega_B=0.04$, and $T=0.3$ in (a), and $T=0.06$ in (b).
}
    \label{fig4}
\end{figure}

We now assume that the temperature $T$ of the thermal bath is known and focus on estimating the magnetic field $B_{\text{ext}}$. We aim to determine whether the quadratic interactions–induced squeezing and non-Gaussianity enhance the precision of magnetic-field estimation compared to the radiation-pressure interaction case. We first plot the QFI for estimating $B_{\text{ext}}$ as a function of $B_{\text{ext}}$ in Fig.~\ref{fig4} for both interaction Hamiltonians.

In Fig.~\ref{fig4}(a), we fix the temperature to $T=0.3$ and examine the effect of different values of the coupling strength $g$ under Eq.~\eqref{mod2}. The QFI reaches its maximum for $g=0.08$. By optimizing the parameters, we find that for $g=0.08$ the maximum value of the QFI is obtained at $T=0.3$. We therefore conclude that, when using nonlinear interaction Hamiltonian~\eqref{mod2}, we achieve higher magnetic-field sensitivity for $g=0.08$ and $T=0.3$.

In Fig.~\ref{fig4}(b), we plot the QFI obtained under the quadratic interaction Hamiltonian~\eqref{mod1}. By carefully selecting the system parameters, we find that the maximum QFI for estimating $B_{\text{ext}}$ is achieved in the low-temperature regime. Accordingly, we fix the temperature at $T=0.06$ and analyze the QFI for different values of the coupling strength $g$.

For $T=0.06$ and $g=0.04$, the QFI is significantly larger than that obtained under the effective interaction Hamiltonian~\eqref{mod2}. This indicates that, when the radiation-pressure interaction is considered, achieving maximum precision in magnetic-field estimation requires operating in a relatively high-temperature regime. In contrast, when the quadratic interaction Hamiltonian~\eqref{mod1} is employed, the highest sensitivity is obtained at low temperatures.

The precision of magnetic-field estimation is enhanced at weak $B_{\text{ext}}$ due to interaction-induced squeezing and moderate non-Gaussianity. In particular, the highest magnetic-field sensitivity is achieved near $g=0.04$, where the probe state remains only weakly non-Gaussian, and genuine quadrature squeezing dominates. As the coupling strength increases further and the state enters the strongly non-Gaussian regime where squeezing disappears, the magnetic-field QFI decreases. This indicates that squeezing is the primary resource for magnetic-field estimation in our scheme, whereas strong non-Gaussianity plays a more prominent role in enhancing temperature sensitivity.
%*******************************************************************************%
%-----------------------------------------------------------------------%
%-----------------------------------------------------------------------%

\subsection{Practical measurement considerations}
In this section, we first analyze the classical information accessible using the full statistics of the probe state and subsequently compare it with the experimentally accessible observables.

\subsubsection{Classical Fisher information from full measurement statistics}
To quantify the parameter sensitivity, we evaluate the CFI associated with the full photon-number $p(n)$ and quadrature probability distributions $p(x)$. These distributions correspond to the measurement statistics obtained through photon-counting and homodyne detection, respectively.

For a generic parameter $\theta \in \{T,B\}$ and an arbitrary measurement distribution $p(\mu|\theta)$, the corresponding CFI is defined as~\cite{helstrom1969quantum,Helstrom1969,paris2009}
\begin{align}
F_{C}(\theta)= \sum_{\mu}
\frac{ \left[ \partial_{\theta} p(\mu|\theta)
\right]^{2} }{ p(\mu|\theta) }.
\end{align}
For photon-counting measurements, the probability distribution is given by
\begin{align} p(n|\theta)
= \langle n|\rho(\theta)|n\rangle,
\end{align}
which corresponds to the photon-number statistics, whereas for the quadrature measurements obtained, it reads as
\begin{align}
p(x|\theta) =
\langle x|\rho(\theta)|x\rangle.
\end{align} 
The quadrature probability distribution $p(x)$ is reconstructed from the reduced density matrix in the Fock basis according to~\cite{walls2008quantum}
\begin{align}
p(x)=\sum_{n,m}\rho_{nm}\,\psi_n(x)\psi_m^{*}(x),
\end{align}
where $\rho_{nm} = \langle n | \rho | m \rangle$ are the density-matrix elements in the Fock basis and $\psi_n(x) = \langle x | n \rangle$ denote the harmonic-oscillator wavefunctions in the position representation. This corresponds to homodyne detection of the field quadrature and enables direct evaluation of the continuous-variable classical Fisher information.

\begin{figure}[t!]
    \centering
    \includegraphics[scale=0.66]{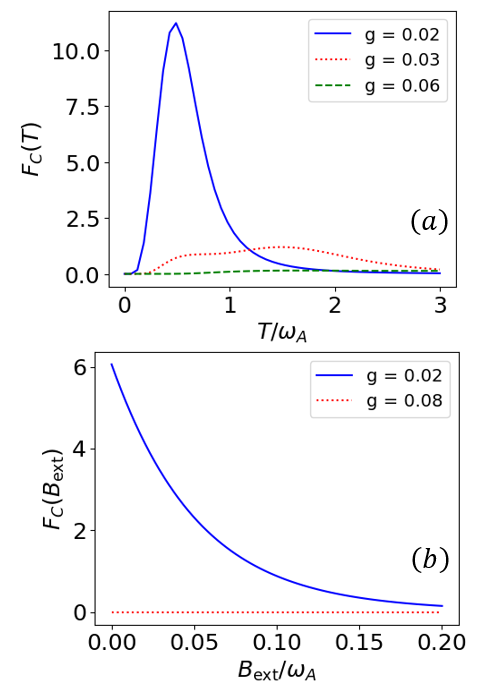}
    \caption{(a) CFI $F_C(T)$ for temperature estimation obtained from photon-counting statistics $p(n|T)$ as a function of temperature $T$ for quadratic interaction. Here, the solid blue, red dotted, and green dashed curves correspond to $g=0.02,0.03,0.06$, respectively. (b) CFI $F_C(B_\mathrm{ext})$ for magnetic field estimation obtained from the full quadrature statistics $p(x)$ for quadratic interaction. Here, the solid blue and red dotted curves correspond to $g=0.02$ and $g=0.08$, respectively. The rest of the parameters are set to $\omega_A=1$, $\omega_B=0.04$, and $B_\text{ext}=0.04$.}
    \label{fig:Photon}
\end{figure}

Figure~\ref{fig:Photon}(a) shows the CFI associated with temperature estimation using the full photon-number statistics $p(n|T)$ for different values of the nonlinear coupling strength $g$ using quadratic interaction. We observe that the thermometric sensitivity strongly depends on the coupling strength. For weak nonlinear coupling ($g=0.02$), the CFI exhibits a pronounced peak at low temperatures, indicating enhanced sensitivity of the photon-number distribution to small thermal variations in this regime. As the coupling strength increases ($g=0.03$ and $g=0.06$), the peak value of the CFI becomes strongly suppressed and broadened, demonstrating the reduced temperature sensitivity encoded in the photon-counting statistics. This behavior indicates that the quadratic interaction progressively washes out the thermal distinguishability between neighboring states, thereby reducing the amount of extractable information about temperature.

In contrast, Fig.~\ref{fig:Photon}(b) presents the CFI for magnetic-field estimation obtained from the full quadrature distribution $p(x|B_{\mathrm{ext}})$ which can be reconstructed through homodyne detection. Unlike temperature, the magnetic-field sensitivity displays qualitatively different behavior. For relatively weak coupling strength ($g=0.02$), the quadrature-based CFI attains its maximum value in the low-field regime and decreases monotonically as the magnetic-field strength increases. This indicates that the quadrature distribution is highly sensitive to small displacement-induced perturbations generated by the external magnetic field. However, when the coupling becomes stronger ($g=0.08$), the Fisher information is substantially suppressed over the entire parameter range, indicating that the non-Gaussian regime is unfavorable for high-precision magnetic-field estimation.
%********************************************************%
%********************************************************%
%--------------------------------------------------------------------%
\begin{figure*}[t!]
    \centering
    \subfloat[]{
    \includegraphics[scale=0.48]{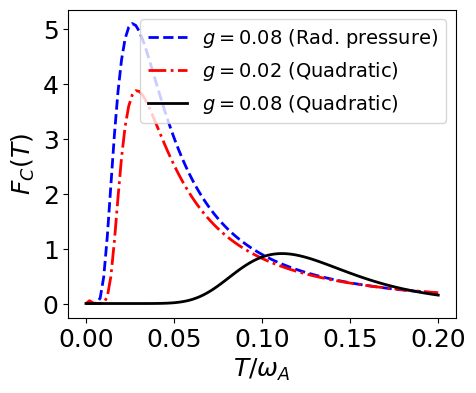}}
    \subfloat[]{
    \includegraphics[scale=0.48]{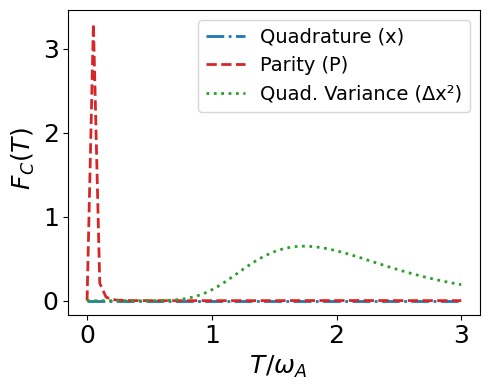}}
    \subfloat[]{
        \includegraphics[scale=0.48]{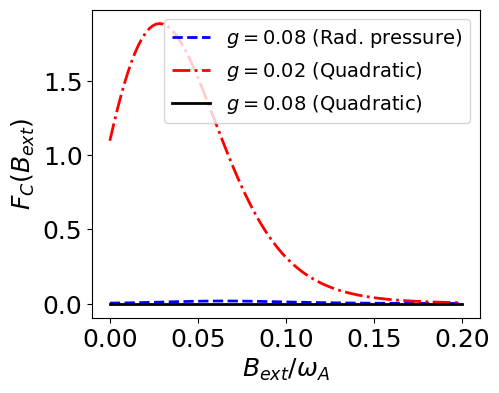}}
    \caption{ (a) $F_C(T)$ as a function of temperature for temperature estimation using mean photon number as an observable. The magnetic field is fixed at $B_{\text{ext}} = 0.04$. Different curves correspond to different coupling strengths $g$ and interaction types. (b) $F_C(T)$ as a function of temperature for different observables under quadratic interaction, with $B_{\text{ext}} = 0.04$. (c) $F_C(B_{\text{ext}})$ as a function of the magnetic field for magnetic field estimation, with the temperature fixed at $T = 0.2$. Different curves correspond to different coupling strengths and interaction types. Unless otherwise specified, the parameters are $\omega_A = 1$ and $\omega_B = 0.04$.}
    \label{fig:CFI}
\end{figure*}
%--------------------------------------------------------------------%
\subsubsection{ Observable-based estimation strategies}
In realistic experimental scenarios, parameter estimation is often constrained by accessible measurement observables rather than by the optimal quantum measurement that saturates the QFI. A natural figure of merit in such cases is the CFI associated with a given measurement scheme. For an observable $\hat{O}$, the sensitivity to a parameter $\lambda$ can be quantified in the error-propagation form
\begin{equation}
F_C(\lambda) = \frac{1}{\mathrm{Var}(\hat{O})} \left( \frac{\partial_{\lambda} \langle \hat{O} \rangle}{\partial \lambda} \right)^2,
\end{equation}
where $\langle \hat{O} \rangle$ and $\mathrm{Var}(\hat{O})$ are the mean and variance of observable $\hat{O}$. In our case, a particularly relevant experimentally accessible observable is the photon number operator $\hat{n} = \hat{a}^\dagger \hat{a}$. Photon-number or intensity-based detection is routinely implemented in optical, circuit quantum electrodynamics, and optomechanical architectures. This demonstrates that the metrological performance analyzed in this work can also be connected to practical measurement protocols available in current quantum sensing platforms.

We plot the CFI as a function of temperature $T$ in Fig.~\ref{fig:CFI}(a) for temperature estimation using the mean photon number as the measurement observable for both interaction models. 
The CFI associated with the radiation-pressure interaction Hamiltonian is consistently larger, and it almost coincides with the QFI. This shows that the mean photon number is the optimal observable for temperature estimation using radiation-pressure interaction. 
However, for the quadratic interaction, the CFI for the same coupling strength ($g=0.08$) is very small compared to radiation-pressure, and the maximum lies in the high temperature regime, while it enhances when we consider a weak coupling strength ($g=0.02$). It is important to note that for weak coupling $g=0.02$, the probe state remains close to Gaussian, and the photon-number statistics are relatively well-behaved. In this regime, photon counting provides significant temperature sensitivity. For stronger coupling $g=0.08$, where non-Gaussianity is maximal, the photon-number distribution develops heavy tails and large fluctuations, and hence the CFI associated with photon-number measurements decreases due to the rapid growth of variance relative to the temperature sensitivity of the mean.

The suppression of the CFI in the strong-coupling quadratic regime indicates that the limitation arises from the choice of measurement we consider. In this regime, the probe becomes strongly non-Gaussian, while the mean photon number probes only the first moment of the photon-number distribution. As a result, temperature-dependent features encoded in higher-order statistics or in the phase-space structure are not efficiently captured by intensity measurements.

To explore this aspect, we consider alternative observables that are sensitive to non-Gaussian features, namely the field quadrature $\hat{x} = (\hat{a}+\hat{a}^\dagger)/\sqrt{2}$, its second moment, and the photon-number parity operator $\hat{\Pi}=(-1)^{\hat{n}}$ whose expectation value $\langle \hat{\Pi} \rangle$ varies smoothly with temperature. We plot the CFI as a function of $T$ in Fig.~\ref{fig:CFI}(b) for these different observables. While quadrature-based measurements provide only limited improvement, parity measurements yield a noticeably larger CFI in both the strong-coupling and low-temperature regimes. This indicates that, in the non-Gaussian regime, temperature information is more effectively encoded in the parity structure of the photon-number distribution than in its first moment.

We further employ the mean photon number for magnetic-field sensing in Fig.~\ref{fig:CFI}(c). For weak coupling ($g=0.02$), the quadratic interaction [cf. Eq.~\eqref{mod1}] significantly outperforms the nonlinear interaction [cf. Eq.~\eqref{mod2}] by enabling a more efficient transduction of magnetic-field variations into changes in the probe’s mean photon number. In this regime, the probe state remains close to Gaussian and exhibits genuine quadrature squeezing, which enhances its susceptibility to small magnetic-field variations and leads to an increased CFI. In contrast, for stronger coupling ($g=0.08$), the probe enters a strongly non-Gaussian regime where squeezing disappears, and both interactions yield a strongly suppressed CFI. This indicates that magnetic-field sensitivity is primarily supported by squeezing at moderate coupling, while the strongly non-Gaussian regime is less effective for magnetometry.
%---------------------------------------------------------%
\begin{figure*}[t!]
    \centering
    \subfloat[]{
    \includegraphics[scale=0.75]{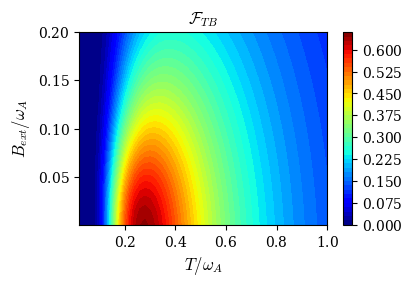}}
        \subfloat[]{
    \includegraphics[scale=0.75]{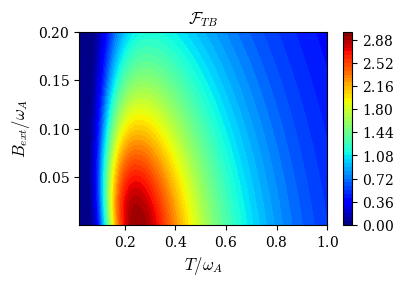}}\\
        \subfloat[]{
    \includegraphics[scale=0.75]{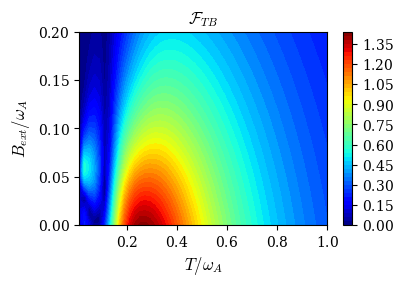}}
        \subfloat[]{
   \includegraphics[scale=0.75]{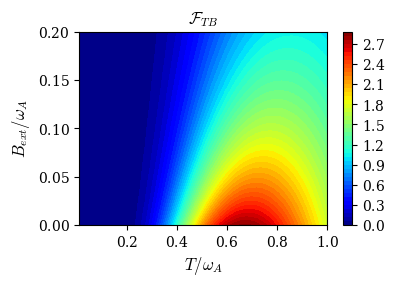}}
    \caption{Top row: Off-diagonal element $\mathcal{F}_{TB_\mathrm{ext}}$ of the QFIM for the joint estimation of bath temperature $T$ and magnetic field $B_{\text{ext}}$, obtained using the radiation-pressure interaction Hamiltonian~\eqref{mod2}. The other parameters are fixed at $\omega_A = 1$ and $\omega_B = 0.04$. Panels (a) and (b) correspond to coupling strengths $g = 0.02$ and $g = 0.08$, respectively. Bottom row: Off-diagonal element $\mathcal{F}_{TB_\mathrm{ext}}$ of the QFIM for the same estimation task, obtained using the quadratic interaction Hamiltonian~\eqref{mod1}. Panels (a) and (b) correspond to $g = 0.02$ and $g = 0.08$, respectively.
}
    \label{fig6}
\end{figure*}
%--------------------------------------------------------%
In particular, the metrological advantage of the quadratic interaction is evident only in the weak-coupling regime, where a fraction of the quantum enhancement can be extracted using simple and experimentally feasible measurements. This demonstrates that, while optimal measurements are necessary to attain the quantum limit, practical observables such as the mean photon number can still provide precision gains when the system operates in the appropriate coupling regime. Interestingly, although the probe state becomes strongly non-Gaussian for larger coupling $g$, this regime does not enhance the CFI when the mean photon number is employed as the measurement observable. This indicates that the non-Gaussian features of the state are not efficiently encoded in the first moment of the photon-number distribution. Consequently, non-Gaussianity alone is not sufficient to guarantee improved metrological performance under restricted measurement strategies. 

Importantly, in both estimation protocols discussed above, the use of the full probability distributions $p(n)$ and $p(x)$ yields larger CFI compared to estimators based on low-order observables such as mean photon number, parity, or quadrature moments. This shows that the photon and homodyne statistics retain additional parameter-dependent information that is lost under moment-based coarse graining. Consequently, full-distribution measurements provide a more informative and experimentally consistent framework for parameter estimation.

\subsection{Joint estimation of temperature and magnetic field}
After analyzing the precision of single-parameter estimation for temperature and magnetic field, we now turn to their joint estimation. For the parameter vector $\boldsymbol{\theta} = (T, B_{\text{ext}})$, the QFI generalizes to the QFIM, which can be written as
\begin{equation}
\mathcal{\boldsymbol{F}}_Q(T,B_{\text{ext}})
=
\begin{pmatrix}
\mathcal{F}_{TT} & \mathcal{F}_{TB_\mathrm{ext}} \\
\mathcal{F}_{B_\mathrm{ext}T} & \mathcal{F}_{B_\mathrm{ext}B_\mathrm{ext}}
\end{pmatrix}.
\end{equation}
The diagonal elements $\mathcal{F}_{TT}$ and $\mathcal{F}_{B_\mathrm{ext}B_\mathrm{ext}}$ quantify the ultimate precision bounds for estimating temperature and magnetic field individually, while the off-diagonal elements $\mathcal{F}_{TB_\mathrm{ext}} = \mathcal{F}_{B_\mathrm{ext}T}$ capture the statistical correlations arising in the simultaneous estimation of both parameters. For two parameters, the determinant
\begin{equation}
\det \mathcal{\boldsymbol{F}}_Q = \mathcal{F}_{TT}\mathcal{F}_{B_\mathrm{ext}B_\mathrm{ext}} - \mathcal{F}_{TB_\mathrm{ext}}^2
\end{equation}
provides a compact measure of the achievable joint estimation precision.

These correlations determine whether joint estimation leads to compatible or competing optimal measurements, thereby constraining the achievable precision through the multiparameter QCRB.
The density matrix $\hat{\rho}_T$ in Eq.~\eqref{gibbs} encodes the parameters of interest for which joint estimation is considered. We numerically compute the QFIM for the probe state, which sets a lower bound on the variance of any unbiased estimator. The single-parameter estimation results, corresponding to the diagonal elements of the QFIM and representing the maximum achievable precision in estimating $T$ and $B_\mathrm{ext}$, are shown in Figs.~\ref{fig2}, \ref{fig3}, and \ref{fig4}. 

In Fig.~\ref{fig6}, we present the off-diagonal elements $\mathcal{F}_{TB_\mathrm{ext}}$ of the QFIM for the joint estimation of $T$ and $B_\mathrm{ext}$. Figures~\ref{fig6}(a) and (b) show $\mathcal{F}_{TB_\mathrm{ext}}$ for the probe state obtained under the radiation-pressure interaction Hamiltonian~\eqref{mod2} at two different coupling strengths $g$, where the QFI is larger for $g=0.08$. In Figs.~\ref{fig6}(c) and (d), the QFI $\mathcal{F}_{TB_\mathrm{ext}}$ is plotted for the probe state generated by the quadratic Hamiltonian~\eqref{mod1}, for which we observe an enhancement compared to the radiation-pressure interaction. In this case, for weak coupling strengths $g$, the off-diagonal element $\mathcal{F}_{TB_\mathrm{ext}}$ attains a relatively larger maximum in the intermediate-temperature regime. As the coupling strength $g$ is increased, the peak value of $\mathcal{F}_{TB_\mathrm{ext}}$ is significantly amplified; however, this enhancement occurs within a much narrower region of the parameter space. We observe that the off-diagonal element $\mathcal{F}_{TB_\mathrm{ext}}$ vanishes in both the low- and high-temperature limits, while attaining its maximum at intermediate temperatures. Increasing the coupling strength $g$ leads to a moderate enhancement of the peak value of $\mathcal{F}_{TB_\mathrm{ext}}$, without changing its qualitative behavior. Furthermore, the statistical correlation between temperature and magnetic field weakens as the external field strength $B_{\mathrm{ext}}$ increases. The relatively small magnitude of $\mathcal{F}_{TB_\mathrm{ext}}$ indicates that the two parameters are only weakly correlated overall, with the strongest statistical correlation occurring in the intermediate-temperature regime.

A comparison between the joint estimation of $\{T,B_{\mathrm{ext}}\}$ and the corresponding single-parameter estimation reveals that, for both interaction models, the QFI obtained in the single-parameter scenario exceeds that achieved under joint estimation. This reduction in precision originates from the non-diagonal structure of the QFIM, indicating statistical correlations between the two parameters, which redistribute the available information. In the quadratic interaction model, the counter-rotating terms generate intrinsic squeezing at intermediate coupling and strong non-Gaussianity at larger coupling strengths. These features enhance the precision of single-parameter estimation; however, they also increase statistical correlations between temperature and magnetic-field estimators in the joint-estimation scenario. Consequently, within the parameter regime considered here, joint estimation does not outperform independent single-parameter strategies, as evidenced by the non-diagonal elements of the QFIM.

In multiparameter quantum estimation, limitations on the attainable precision may arise either from statistical correlations between the estimated parameters or from the incompatibility of the optimal measurements. These effects can be distinguished by analyzing the SLDs associated with each parameter.

For the joint estimation of the temperature $T$ and the magnetic field $B_{\mathrm{ext}}$, the off-diagonal elements of the QFIM quantify statistical correlations between the parameters but do not, by themselves, determine whether the corresponding optimal measurements are compatible. To directly assess measurement compatibility associated with the optimal observables, we evaluate the expectation value of the commutator of the SLDs corresponding to $T$ and $B_{\mathrm{ext}}$ using~\cite{PhysRevA.94.052108,Matsumoto_2002}
\begin{equation}
\mathcal{C}_{TB_\mathrm{ext}} \equiv \frac{1}{2i}\,\left\{\mathrm{Tr}\!\left[\rho\,(L_T L_{B_{\mathrm{ext}}} - L_{B_{\mathrm{ext}}} L_T)\right]\right\},
\label{eq:SLD_comm}
\end{equation}
which is commonly referred to as the \emph{sufficient commutation condition}. A vanishing value of $\mathcal{C}_{TB_{\mathrm{ext}}}$ implies that the QCRB for multiparameter estimation can, in principle, be saturated.

For completeness, we also consider the real part of the expectation value of the SLDs, such as
\begin{equation}
\mathcal{R}_{TB_{\mathrm{ext}}} \equiv \mathrm{Re}\!\left\{\mathrm{Tr}\!\left[\rho\,(L_T L_{B_{\mathrm{ext}}} - L_{B_{\mathrm{ext}}} L_T)\right]\right\}.
\end{equation}
We evaluate both quantities to provide a complete characterization of the expectation value of the SLD commutator. For both the radiation-pressure and quadratic interaction Hamiltonians, we find numerically that the imaginary part $\mathcal{C}_{TB_\mathrm{ext}}$ remains negligibly small across the entire parameter regime considered. This indicates that the optimal SLDs associated with temperature and magnetic field estimation satisfy the commutation condition; therefore, the corresponding optimal measurements are compatible. The real part $\mathcal{R}_{TB_\mathrm{ext}}$ also remains several orders of magnitude smaller than the relevant QFI scales and exhibits no systematic structure across the parameter space. As a result, the reduced precision observed in the joint estimation of $T$ and $B_{\mathrm{ext}}$ compared to single-parameter estimation cannot be attributed to measurement incompatibility. Instead, it originates from statistical correlations encoded in the off-diagonal elements of the QFIM. Hence, although the quadratic interaction induces squeezing and non-Gaussianity that enhances single-parameter sensitivity, it does not induce incompatibility between the optimal measurements for $T$ and $B_{\mathrm{ext}}$.
%************************************************************************%

\section{Conclusion}\label{conc}
We investigated single- and multiparameter quantum metrology using a coupled resonator system, focusing on two different interaction models. By employing the QFI as a figure of merit, we examined how the form of the interaction influences the estimation precision of temperature and magnetic field. Our results demonstrate that the quadratic interaction between resonator~A and resonator~B, described by the interaction Hamiltonian~\eqref{mod1}, provides a substantial enhancement in the estimation precision of parameters compared to the radiation-pressure interaction Hamiltonian~\eqref{mod2} due to the interaction-induced squeezing at intermediate coupling and strong non-Gaussian features at stronger coupling strengths. Notably, our scheme does not rely on external driving or conditional measurements; rather, the non-Gaussian probe state emerges intrinsically from the counter-rotating and quadratic interaction terms, making the metrological enhancement an inherent property of the system.

The improvement in estimation precision can be understood by analyzing the Wigner functions of resonator~A generated under the two interaction Hamiltonians. We first computed the Wigner function for the reduced state of resonator~A for both interaction models. For the quadratic interaction Hamiltonian~\eqref{mod1}, we observe squeezing in the momentum quadrature of the Wigner function at intermediate coupling. This squeezing dominates in the intermediate-coupling regime, whereas at stronger coupling it disappears and the probe enters a strongly non-Gaussian regime. In the strong coupling regime, the Wigner function exhibits two well-separated coherent lobes accompanied by pronounced interference fringes and negative regions. We then quantify the non-Gaussian features of the probe state using non-Gaussian measures and kurtosis, thereby further confirming its non-Gaussian nature. For the same set of parameters, our analysis of the QFI shows that the enhancement in the precision of temperature and magnetic-field estimation is associated with this interaction-induced squeezing in the intermediate-coupling regime and from strong non-Gaussianity in the strong-coupling regimes. In particular, the non-Gaussianity of the probe state redistributes fluctuations and leads to a substantial increase in the QFI in the low-temperature regime, resulting in improved temperature and magnetic-field estimation precision.

We also investigated the multiparameter estimation problem, namely the joint estimation of temperature and magnetic field, and analyzed the behavior of the QFIM as a function of both parameters. We find that the precision achievable in the joint estimation scenario is reduced compared to single-parameter estimation. This reduction originates from statistical correlations encoded in the off-diagonal elements of the QFIM, which become more pronounced in the strongly non-Gaussian regime. Importantly, despite the presence of these correlations, the expectation value of the commutator of the SLDs remains negligible, indicating that the optimal measurements for temperature and magnetic field are compatible. Therefore, the degradation in joint-estimation precision cannot be attributed to measurement incompatibility but rather to the redistribution of information among correlated parameters. Our results indicate that, for the coupled resonator system considered here, single-parameter estimation remains the most effective strategy for high-precision thermometry and magnetometry.

\section*{Acknowledgment}
This work was supported by the Scientific and Technological Research Council of Türkiye (TÜBİTAK) under Project Grant Nos. 122F371 and 123F150.
\section*{DATA AVAILABILITY}
The data that support the findings of this article are not publicly available. The data are available from the authors upon reasonable request.
\appendix
\section{Robustness against higher-order nonlinear corrections}\label{Robustness}
To assess the robustness of our results against higher-order nonlinearities that may arise in specific physical implementations, we consider an extended Hamiltonian obtained by supplementing the minimal model with representative quartic self-interaction and cross-quartic nonlinear terms,
\begin{equation}
\hat{H}_{\mathrm{ext}}
=
\hat{H}_{\mathrm{min}}
+\kappa_{4}(\hat{a}+\hat{a}^{\dagger})^{4}
+\kappa_{22}(\hat{a}+\hat{a}^{\dagger})^{2}
(\hat{b}+\hat{b}^{\dagger})^{2},
\label{eq:Hext}
\end{equation}
where $\kappa_{4}$ and $\kappa_{22}$ denote the coupling strengths of quartic self-interaction and cross quartic nonlinear interactions, while \(\hat{H}_{\mathrm{min}}\) denotes the minimal effective Hamiltonian introduced in the main text. The quartic self-interaction term represents a generic higher-order nonlinear correction to the effective bosonic mode, while the mixed quartic term models higher-order cross-mode nonlinearities that may emerge in microscopic superconducting-circuit, optomechanical, or related hybrid quantum implementations. These additional terms are not assumed to be universal. Rather, they serve as representative higher-order nonlinear corrections that may arise in particular microscopic realizations of the effective model.
%-------------------------------------------------------------%
\begin{figure*}
    \centering
    \subfloat[]{
    \includegraphics[width=0.33\linewidth]{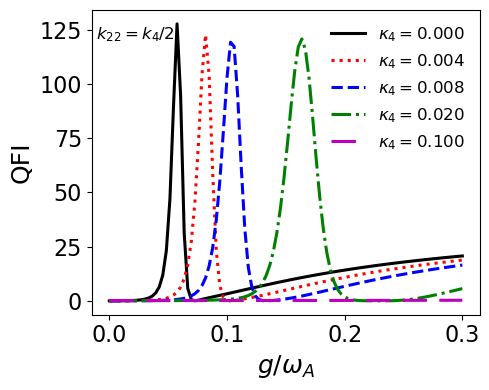}}
        \subfloat[]{
    \includegraphics[width=0.33\linewidth]{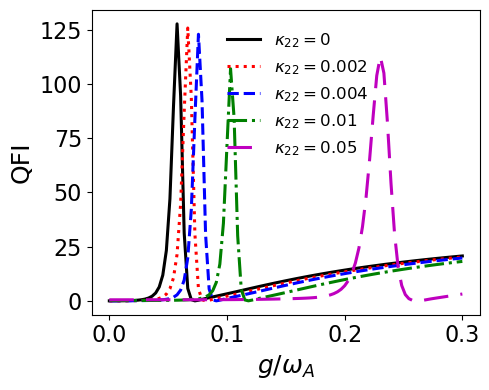}}
        \subfloat[]{
        \includegraphics[width=0.33\linewidth]{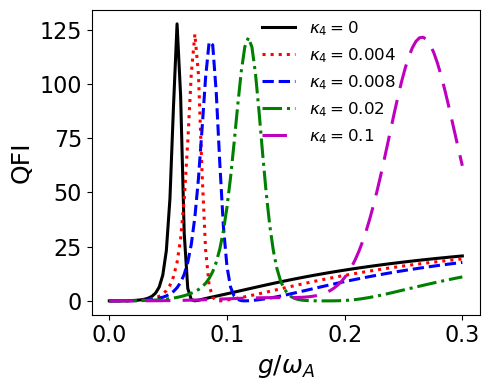}}
    \caption{The QFI as a function of the coupling strength $g$ for temperature estimation using the extended Hamiltonian in Eq.~\eqref{eq:Hext}. (a) Both higher-order nonlinear coefficients are included, with $\kappa_{22}=\kappa_4/2$, and $\kappa_4=0$, $0.004$, $0.008$, $0.020$, and $0.100$. (b) The quartic nonlinear coefficient is absent ($\kappa_4=0$), while $\kappa_{22}=0$, $0.002$, $0.004$, $0.010$, and $0.050$. (c) The mixed nonlinear coefficient is absent ($\kappa_{22}=0$), while $\kappa_4=0$, $0.004$, $0.008$, $0.020$, and $0.100$. Unless otherwise stated, the remaining parameters are fixed at $\omega_A=1$, $\omega_B=0.04$, $T=0.06$, $B_{\rm ext}=0.04$, $\kappa_4=0.008$, and $\kappa_{22}=0.004$. }
    \label{fig:H_ext}
\end{figure*}
%-------------------------------------------------------------%
To assess the robustness of our results against implementation-dependent higher-order nonlinearities, we systematically investigate the extended Hamiltonian~\eqref{eq:Hext} by independently and jointly varying the quartic self-interaction coefficient $\kappa_4$ and the mixed quartic coefficient $\kappa_{22}$. Figure~\ref{fig:H_ext}(a) considers the representative parameter sets $(\kappa_{4},\kappa_{22})=(0,0)$, $(0.004,0.002)$, $(0.008,0.004)$, $(0.020,0.010)$, and $(0.100,0.050)$, satisfying the relation $\kappa_{22}=\kappa_4/2$. Figures~\ref{fig:H_ext}(b) and \ref{fig:H_ext}(c) isolate the effects of the mixed and quartic nonlinearities by varying $\kappa_{22}$ with $\kappa_4=0$ and varying $\kappa_4$ with $\kappa_{22}=0$, respectively.

The results demonstrate that the principal metrological enhancement predicted by the minimal effective model is robust against moderate higher-order corrections. In all three cases, weak and intermediate nonlinear corrections leave the maximum attainable QFI nearly unchanged while shifting the optimal operating point toward larger values of the effective coupling $g$. This indicates that the dominant effect of higher-order nonlinearities is to renormalize the optimal sensing regime rather than degrade the achievable estimation precision.

A comparison of Figs.~\ref{fig:H_ext}(b) and \ref{fig:H_ext}(c) reveals distinct roles of the mixed quartic and quartic self-interaction nonlinearities. When only the mixed nonlinear coefficient $\kappa_{22}$ is varied [$\kappa_4=0$, Fig.~\ref{fig:H_ext}(b)], the QFI peak shifts gradually toward larger coupling strengths as $\kappa_{22}$ increases, while its maximum value changes slightly over weak and intermediate nonlinearities. A significant modification appears only for the largest value, $\kappa_{22}=0.05$, where the optimal operating point moves substantially toward stronger coupling. In contrast, when only the quartic self-interaction is present [$\kappa_{22}=0$, Fig.~\ref{fig:H_ext}(c)], increasing $\kappa_4$ produces a more rapid displacement of the QFI peak toward higher coupling strengths. Moreover, for sufficiently large quartic nonlinearity ($\kappa_4=0.1$), the maximum of QFI shifts towards much stronger coupling strength. These results indicate that the quartic self-interaction modifies the metrological landscape more efficiently than the mixed quartic interaction, particularly in the strong-nonlinearity regime.

Only when the higher-order nonlinearities become sufficiently large, for example
$(\kappa_{4},\kappa_{22})=(0.100,0.050)$, does the sharp QFI resonance disappear, and the attainable precision becomes reduced throughout the parameter regime considered (as presented in Fig.~\ref{fig:H_ext}(a)). In the case of individual nonlinear terms, our results demonstrate that the higher-order corrections remain perturbative over a broad parameter range and primarily shift the optimal operating point, while the qualitative metrological advantages of the minimal effective model are preserved (as given in Figs.~\ref{fig:H_ext}(b) and \ref{fig:H_ext}(c)). Significant degradation occurs only when the higher-order nonlinearities become comparable to the dominant nonlinear interaction, beyond the regime in which the effective low-energy description is expected to remain quantitatively accurate.

\begin{figure}[b!]
    \centering
    \includegraphics[width=\linewidth]{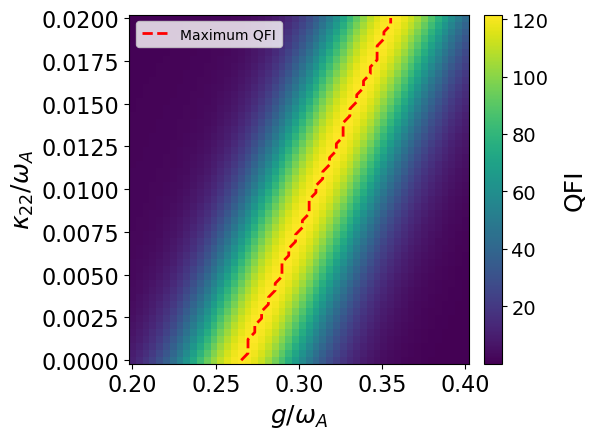}
    \caption{QFI for temperature estimation as a function of coupling strength $g$ and the higher-order nonlinear coefficient $\kappa_{22}$ for a fixed quartic nonlinear coefficient $\kappa_{4}=0.1$. The white dashed line denotes the optimal interaction strength that maximizes the QFI for each value of $\kappa_{22}$. Other parameters are the same as in Fig.~\ref{fig:H_ext}.}
    \label{fig:H_ext_heatmap}
\end{figure}

While Figs.~\ref{fig:H_ext}(b) and \ref{fig:H_ext}(c) reveal the individual influence of the mixed quartic and quartic self-interaction nonlinearities, practical implementations generally involve the simultaneous presence of both higher-order corrections. In this case, the optimal metrological performance is determined by the combined interplay among the interaction strength $g$, the mixed nonlinear coefficient $\kappa_{22}$, and the quartic self-interaction coefficient $\kappa_{4}$. Therefore, it is instructive to identify the optimal operating region in this multidimensional parameter space. To this end, Fig.~\ref{fig:H_ext_heatmap} presents a two-dimensional QFI map as a function of $g$ and $\kappa_{22}$ for a representative value $\kappa_{4}=0.1$, illustrating how the interaction parameters should be jointly optimized to maximize the attainable estimation precision. Figure~\ref{fig:H_ext_heatmap} shows that the highest QFI is confined to a narrow ridge in the $(g,\kappa_{22})$ parameter space, rather than occurring at an isolated value of either parameter. This behavior reflects the coexistence of multiple higher-order nonlinear interaction channels, whose collective influence on the phase space prevents any single nonlinear term from solely determining the metrological performance. The red dashed curve in Fig.~\ref{fig:H_ext_heatmap} traces the optimal interaction  $g$ that maximizes the QFI for each value of $\kappa_{22}$, demonstrating that the optimal coupling shifts toward larger values as the mixed nonlinear coefficient increases. These results indicate that the metrological enhancement is governed by the combined interplay between $g$, $\kappa_{22}$, and $\kappa_{4}$, and that achieving the highest temperature-estimation precision requires joint optimization of these nonlinear parameters rather than tuning the interaction strength alone.
%****************************************************************%
%****************************************************************%
\section{Derivation of the quadratic interaction Hamiltonian (Eq.~\eqref{mod1})}\label{Eq.(4)}
In this appendix, we briefly summarize the derivation of the effective optomechanical interaction Hamiltonians used in the main text, following the canonical moving-boundary formulation introduced by Law~\cite{PhysRevA.51.2537}. We focus on the physical assumptions and approximations leading to Eqs.~\eqref{mod1} and~\eqref{mod2}, namely the single-mode approximation and the rotating-wave approximation (RWA).

Let us consider a one-dimensional Fabry-P\'{e}rot cavity with a fixed mirror at $x=0$ and a movable mirror at $x=q(t)$ of mass $m$ held by potential $V(q)$.
The cavity field $A(x,t)$ satisfies the wave equation (with $c = 1$):
\begin{equation}
  \frac{\partial^{2} A}{\partial x^{2}}
  \;=\;
  \frac{\partial^{2} A}{\partial t^{2}}
  \label{eq:wave}
\end{equation}
with Dirichlet boundary conditions at both surfaces such as $A(0,t) = 0$ and $A\bigl(q(t),\,t\bigr) = 0$.

The cavity field is in instantaneous normal modes
that satisfy the moving boundary condition at every instant:
\begin{equation}
  A(x,t) = \sum_{k=1}^{\infty} Q_{k}(t)\,
  \sqrt{\frac{2}{q(t)}}\,
  \sin\!\left(\frac{k\pi x}{q(t)}\right),
  \label{eq:app_Aexpand}
\end{equation}
where $Q_{k}(t)$ is the amplitude of mode $k$ and the
instantaneous resonance frequency is $\omega_{k}(q)=k\pi/q$. Because the mode functions depend on $q(t)$, taking a time derivative of $Q_{k}$ mixes different modes. The mixing strength between mode $k$ and mode $j$ is given by the purely geometric numbers~\cite{PhysRevA.51.2537}
\begin{equation}
  g_{kj}
  \;=\;
  \begin{cases}
    \displaystyle
    (-1)^{k+j}\,\frac{2kj}{j^{2}-k^{2}}
    & k \neq j, \\[8pt]
    0 & k = j.
  \end{cases}
  \label{eq:gkj}
\end{equation}
which are antisymmetric ($g_{kj}=-g_{jk}$) and vanish when the mirror is stationary.

Substituting the instantaneous-mode expansion~\eqref{eq:app_Aexpand} into the wave equation and the mirror equation of motion yields the coupled field-mirror dynamics derived in Ref.~\cite{PhysRevA.51.2537}
\begin{align}
  \ddot{Q}_{k} =& -\omega_{k}^{2}(q)Q_{k} + \frac{2\dot{q}}{q}\sum_{j} g_{kj}\dot{Q}_{j} + \frac{\ddot{q}q - \dot{q}^{2}}{q^{2}}\sum_{j} g_{kj}Q_{j} \\
  +& \frac{\dot{q}^{2}}{q^{2}}\sum_{j,l} g_{jk}g_{jl}Q_{l}, \label{eq:Qddot} \\
  m\ddot{q} =& -\frac{\partial V}{\partial q} + \frac{1}{q}\sum_{k,j}(-1)^{k+j}\omega_{k}\omega_{j}Q_{k}Q_{j}. \label{eq:qddot}
\end{align}
The second term in Eq.~\eqref{eq:qddot} represents the radiation pressure force exerted by the electromagnetic field on the moving boundary. As shown in Ref.~\cite{PhysRevA.51.2537}, this coupled system is exactly described by the Lagrangian
\begin{align}
  L = \frac{1}{2}\sum_{k}\bigl[\dot{Q}_{k}^{2} - \omega_{k}^{2}(q)Q_{k}^{2}\bigr] + \frac{1}{2}m\dot{q}^{2} - V(q) \\
  - \frac{\dot{q}}{q}\sum_{k,j} g_{kj}\dot{Q}_{k}Q_{j} + \frac{\dot{q}^{2}}{2q^{2}}\sum_{k,j,l} g_{kj}g_{kl}Q_{j}Q_{l}. \label{eq:Lag}
\end{align}
The canonical momenta associated with $Q_k$ and $q$ are
\begin{equation}
  P_{k} = \dot{Q}_{k} - \frac{\dot{q}}{q}\sum_{j} g_{kj}Q_{j}, \quad \text{and} \quad p = m\dot{q} - \frac{1}{q}\sum_{k,j} g_{kj}P_{k}Q_{j}.
\end{equation}
Note that $p$ differs from the kinetic momentum $m\dot{q}$ by a field-dependent term, analogous to the minimal coupling $p = m\dot{x} + eA$ in electrodynamics. Performing a Legendre transformation $H = p\dot{q} + \sum_{k}P_{k}\dot{Q}_{k} - L$, we obtain the classical Hamiltonian:
\begin{align}
  H = \frac{1}{2m}\left( p + \frac{1}{q}\sum_{k,j} g_{kj}P_{k}Q_{j} \right)^{2} + V(q) \\
  + \frac{1}{2}\sum_{k}\bigl[P_{k}^{2} + \omega_{k}^{2}(q)Q_{k}^{2}\bigr]. \label{eq:Hclass}
\end{align}
This expression represents the total energy of the coupled system, partitioned into the mirror's kinetic energy, the mechanical potential, and the field energy of the instantaneous cavity modes.
We replace all classical dynamical variables by operators satisfying $[\hat{q},\hat{p}]=i$,
$[\hat{Q}_{k},\hat{P}_{j}]=i\delta_{kj}$
and and introduce ladder operators $\hat{a}_{k}$, $\hat{a}_{k}^{\dagger}$ for each cavity mode via
\begin{align}
  \hat{Q}_{k}
  &= \frac{1}{\sqrt{2\omega_{k}(q)}}
     \left(\hat{a}_{k} + \hat{a}_{k}^{\dagger}\right),
  \label{eq:Qinv}\\[6pt]
  \hat{P}_{k}
  &= i\sqrt{\frac{\omega_{k}(q)}{2}}
     \left(\hat{a}_{k}^{\dagger} - \hat{a}_{k}\right).
  \label{eq:Pinv}
\end{align}
Substituting Eqs.\,\eqref{eq:Qinv}--\eqref{eq:Pinv} into
the classical Hamiltonian~\eqref{eq:Hclass}, and replacing the divergent vacuum sum with the Casimir energy $-\pi/(24q)$ (standard procedure in 1D), gives the full quantum Hamiltonian
\begin{equation}
  \hat{H}
  = \frac{(\hat{p}+ \hat{\Gamma})^{2}}{2m}
  + V(\hat{q})
  + \sum_{k} \omega_{k}(\hat{q})\,\hat{a}_k^\dagger\,\hat{a}_k
  - \frac{\pi}{24\hat{q}},
  \label{eq:Hquant}
\end{equation}
where the operator $\hat{\Gamma}$ encodes the coupling between mirror motion and field modes
\begin{equation}
  \hat{\Gamma}
  = \frac{i}{2\hat{q}}
    \sum_{k,j} g_{kj}\sqrt{\frac{k}{j}}\;
    \Bigl[
      \hat{a}_k^\dagger\hat{a}_{j}^{\dagger}
      - \hat{a}_k\hat{a}_{j}
      + \hat{a}_k^\dagger\hat{a}_{j}
      - \hat{a}_{j}^{\dagger}\hat{a}_k
    \Bigr].
  \label{eq:Gamma}
\end{equation}
To obtain the effective low-energy optomechanical Hamiltonian, several controlled approximations are introduced. First, the mirror displacement is assumed to remain small compared to the equilibrium cavity length, $|\hat{x}| \ll q_0$, allowing an expansion around the equilibrium configuration. Second, only a single near-resonant cavity mode is retained, while intermode scattering into other optical modes is neglected under the adiabatic single-mode condition discussed below. Finally, rapidly oscillating counter-rotating contributions are neglected within the rotating-wave approximation.
Equation~\eqref{eq:Gamma} contains both \textit{rotating} terms ($\hat{a}_{k}^{\dagger}\hat{a}_{j}$) that conserve the total excitation number, and \textit{counter-rotating} terms ($\hat{a}_{k}^{\dagger}\hat{a}_{j}^{\dagger}$ and $\hat{a}_{k}\hat{a}_{j}$) that create or annihilate pairs of excitations. Expanding Eq.~\eqref{eq:Hquant} around the mirror equilibrium position $q_{0}$ (displacement $\hat{x}=\hat{q}-q_{0}\ll q_{0}$) and applying
a unitary transformation that removes the mixed term
$(\hat{p}+\hat{\Gamma})^{2}/2m$, the Hamiltonian linearizes to
\begin{equation}
  \hat{H}' = \frac{\hat{p}^{2}}{2m} + u(\hat{x})
            + \sum_{k}\omega_{k0}\,\hat{a}_{k}^{\dagger}\hat{a}_{k}
            - \hat{x}\,\hat{F}_{0},
  \label{eq:app_Hlin}
\end{equation}
where $\omega_{k0}=k\pi/q_{0}$ and the radiation pressure
force operator at equilibrium is
\begin{equation}
  \hat{F}_{0}
  = \frac{1}{2q_{0}}\sum_{k,j}(-1)^{k+j}
    \sqrt{\omega_{k0}\omega_{j0}}
    \bigl(
      \hat{a}_{k}\hat{a}_{j}
      +\hat{a}_{k}^{\dagger}\hat{a}_{j}^{\dagger}
      +\hat{a}_{k}^{\dagger}\hat{a}_{j}
      +\hat{a}_{j}\hat{a}_{k}^{\dagger}
    \bigr).
  \label{eq:app_F0}
\end{equation}
Under the single-mode approximation, the force operator in Eq.~\eqref{eq:app_F0} reduces to
\begin{equation}
  \hat{x}\,\hat{F}_{0}
  \approx \hat{x}\,\frac{\omega_{A}}{q_{0}}
  \bigl(
    \hat{a}^{2}+(\hat{a}^{\dagger})^{2}
    +2\hat{a}^{\dagger}\hat{a}+1
  \bigr).
  \label{eq:app_xF}
\end{equation}
Using the identity $(\hat a+\hat a^\dagger)^2 = \hat a^2+(\hat a^\dagger)^2+2\hat a^\dagger\hat a+1$, so that Eq.~\eqref{eq:app_xF} becomes $\hat{x}\hat{F}_{0}=(\omega_{A}/q_{0})\hat{x}(\hat{a}+\hat{a}^{\dagger})^{2}$. Upon quantizing the mirror displacement as resonator~B with frequency $\omega_{B}$ and ladder operators $\hat{b}$, $\hat{b}^{\dagger}$
\begin{equation}
  \hat{x}
  = \frac{1}{\sqrt{2\omega_{B}}}\,(\hat{b}+\hat{b}^{\dagger}),
  \label{eq:app_xquant}
\end{equation} 
and defining the single-photon coupling
$g=\omega_{A}/(q_{0}\sqrt{2\omega_{B}})$, the interaction term becomes $g(\hat{a}+\hat{a}^{\dagger})^{2}(\hat{b}+\hat{b}^{\dagger})$. Adding the free resonator Hamiltonians gives us
\begin{equation}
  \hat{H}
  = \omega_{A}\hat{a}^{\dagger}\hat{a}
  + \omega_{B}\hat{b}^{\dagger}\hat{b}
  + g(\hat{a}+\hat{a}^{\dagger})^{2}(\hat{b}+\hat{b}^{\dagger}),
  \label{eq:app_H4}
\end{equation}
which is Eq.~\eqref{mod1} of the main text which is also used in Ref.~\cite{PhysRevLett.132.053601} to study the quantum phase transitions.

Expanding $(\hat{a}+\hat{a}^{\dagger})^{2}$, the interaction in Eq.~\eqref{eq:app_H4} splits into two parts
\begin{align}
  g(\hat{a}+\hat{a}^{\dagger})^{2}(\hat{b}+\hat{b}^{\dagger})
  = \underbrace{2g\,\hat{a}^{\dagger}\hat{a}\,
    (\hat{b}+\hat{b}^{\dagger})}_{\text{radiation pressure}}\\
  + \underbrace{g\,[\hat{a}^{2}+(\hat{a}^{\dagger})^{2}]
    (\hat{b}+\hat{b}^{\dagger})}_{\text{counter-rotating}}.
  \label{eq:app_split}
\end{align}
Note that a constant displacement of resonator~B has been removed by a shift $\hat{b}\to\hat{b}-g/\omega_{B}$. In the interaction picture defined by the free Hamiltonian, the counter-rotating terms acquire fast time-dependent phases $e^{\pm i(2\omega_{A}\pm\omega_{B})t}$, oscillating at frequency $\approx 2\omega_{A}$. By contrast, the radiation-pressure terms involve $\hat{a}^{\dagger}\hat{a}$, which is time-independent, multiplied by $e^{\pm i\omega_{B}t}$, which is slow. The RWA drops all fast-oscillating terms, valid when the interaction strength is much smaller than the characteristic optical frequency scale, $g \ll \omega_A$. After applying the RWA, only the radiation-pressure term survives, and the Hamiltonian reduces to
\begin{equation}
  \hat{H}_{\mathrm{eff}}
  = \omega_{A}\hat{a}^{\dagger}\hat{a}
  + \omega_{B}\hat{b}^{\dagger}\hat{b}
  + g\,\hat{a}^{\dagger}\hat{a}\,(\hat{b}+\hat{b}^{\dagger}),
  \label{eq:app_H5}
\end{equation}
which corresponds to Eq.~\eqref{mod1} of the main text. In our dimensionless units ($\omega_{A}=1$), the coupling values
$g\in[0.02,0.08]$ satisfy $g\ll 2\omega_{A}=2$, so the RWA is well controlled throughout the parameter range studied.

%**************************************************%
\subsection{Validity of single-mode approximation}
We now briefly summarize the conditions under which the single-mode approximation employed above remains valid. The single-mode approximation is valid when the mechanical motion is adiabatic with respect to the optical mode spacing~\cite{PhysRevA.51.2537,bowen2015quantum}. Let \(A\) denotes the
selected optical mode and let
\begin{equation}
    \Delta_{\rm opt}=\min_{\mu\neq A}|\omega_\mu-\omega_A|
\end{equation}
be the detuning from the nearest discarded optical mode. We require
\begin{equation}
    \omega_B\ll \Delta_{\rm opt},
\end{equation}
where \(\omega_B\) is the mechanical frequency. This condition ensures that
the mechanical sidebands of the retained optical mode remain far from
neighboring cavity resonances, so that photon scattering from the selected
mode into other optical modes is off-resonant. In the moving-boundary
instantaneous-mode formulation, the residual nonadiabatic coupling between
the retained mode \(A\) and a discarded optical mode \(\mu\) scales as
\(C_{\mu A}\dot q/q_0\). Here \(q_0\) is the equilibrium cavity length,
\(q(t)=q_0+x(t)\), and \(C_{\mu A}\) is the dimensionless geometric
mode-mixing coefficient arising from the dependence of the optical mode
functions on the moving boundary. For approximately harmonic mechanical
motion, \(x(t)\sim x_{\rm typ}\cos(\omega_B t)\), where \(x_{\rm typ}\) is the
characteristic mechanical displacement sampled during the dynamics, e.g. the
zero-point displacement for near-ground-state motion or the rms/maximum
displacement for a thermally or coherently excited oscillator. Thus
\(\dot q\sim \omega_B x_{\rm typ}\), and the perturbative amplitude for
off-resonant scattering into mode \(\mu\) is estimated as
\begin{equation}
\left|C_{\mu A}\right|
\frac{\omega_B x_{\rm typ}}
     {q_0|\omega_\mu-\omega_A|}
\ll 1.
\end{equation}
This inequality is the quantitative version of the statement that intermode scattering is negligible. It is distinct from the rotating-wave approximation: the single-mode condition suppresses scattering into other optical modes, whereas the RWA suppresses counter-rotating terms within the retained mode.
\bibliography{OQS.bib}
\end{document}